\documentclass[lettersize,journal]{IEEEtran} 

\usepackage{url}
\usepackage{stfloats}

\usepackage{algorithm, algorithmic,float} 
\usepackage{lipsum}

\usepackage{cite}
\usepackage{xcolor}

\usepackage{amsmath,amssymb,amsfonts,mathrsfs,bm} 
\usepackage{graphicx}   
\usepackage{amsthm} 
\usepackage{caption}
\usepackage{subfigure}
\usepackage{enumerate}

\usepackage{verbatim}  
\usepackage{array} 

\usepackage{multirow}

\usepackage{hyperref}
\hypersetup{
colorlinks=true,
linkcolor=black,
anchorcolor=black,
filecolor=black,
urlcolor=black,
citecolor=black,}

\usepackage[capitalize]{cleveref} 

\usepackage{booktabs} 
\usepackage{makecell}

\newtheorem{theorem}{Theorem}

\newtheorem{lemma}{Lemma}

\newtheorem{assumption}{Assumption}

\newcommand{\Real}{\mathbb{R}}

\newcommand{\Complex}{\mathbb{C}}

\newcommand{\calE}{\mathcal{E}}

\newcommand{\calG}{\mathcal{G}}

\newcommand{\calI}{\mathcal{I}}
\newcommand{\calJ}{\mathcal{J}}

\newcommand{\calS}{\mathcal{S}}
\newcommand{\calT}{\mathcal{T}}

\newcommand{\calV}{\mathcal{V}}

\newcommand{\bA}{\mathbf{A}}

\newcommand{\bD}{\mathbf{D}}
\newcommand{\be}{\mathbf{e}}
\newcommand{\bE}{\mathbf{E}}

\newcommand{\bF}{\mathbf{F}}

\newcommand{\bI}{\mathbf{I}}

\newcommand{\bL}{\mathbf{L}}

\newcommand{\bQ}{\mathbf{Q}}

\newcommand{\bR}{\mathbf{R}}

\newcommand{\bU}{\mathbf{U}}

\newcommand{\bV}{\mathbf{V}}

\newcommand{\bW}{\mathbf{W}}

\newcommand{\bX}{\mathbf{X}}

\newcommand{\bY}{\mathbf{Y}}

\newcommand{\scP}{\mathscr{P}}

\newcommand{\bSigma	}{\bm{\Sigma}}

\begin{document}

\title{Subset Random Sampling and Reconstruction of Finite Time-Vertex Graph Signals}

\author{Hang Sheng \IEEEmembership{Student Member, IEEE}, Qinji Shu, Hui Feng, \IEEEmembership{Member, IEEE}, and Bo Hu, \IEEEmembership{Member, IEEE}

\thanks{This work was supported in part by the National Key R$\&$D Program of China under Grant 2024YFE0200700. (\textit{Corresponding author: Hui Feng.})}

\thanks{H. Sheng, Q. Shu, H. Feng, and B. Hu are with College of Future Information Technology, Fudan University, Shanghai 200433, China (e-mail: hsheng20@fudan.edu.cn, qjshu22@m.fudan.edu.cn, \{hfeng, bohu\}@fudan.edu.cn).
}
\thanks{H. Feng and B. Hu are also with State Key Laboratory of Integrated Chips and Systems, Fudan University, Shanghai 200433, China.}

}

\maketitle

\begin{abstract}
Finite time-vertex graph signals (FTVGS) provide an efficient representation for capturing spatio-temporal correlations across multiple data sources on irregular structures. Although sampling and reconstruction of FTVGS with known spectral support have been extensively studied, the case of unknown spectral support requires further investigation. Existing random sampling methods may extract samples from any vertex at any time, but such strategies are not friendly in practice, where sampling is typically limited to a subset of vertices and moments. To address this requirement, we propose a subset random sampling scheme for FTVGS. Specifically, we first randomly select a subset of rows and columns to form a submatrix, followed by random sampling within that submatrix. In theory, we provide sufficient conditions for reconstructing the original FTVGS with high probability. Additionally, we introduce a reconstruction framework incorporating low-rank, sparsity, and smoothness priors (LSSP), and verify the feasibility of the reconstruction and the effectiveness of the framework through experiments.

\end{abstract}


\section{Introduction}
\label{sec:introduction}

\IEEEPARstart{R}{eal-world} data often exhibits a non-regular structure and is time-varying, as seen in sensor networks \cite{sensor2019,sensor2023}, brain networks \cite{brain2023,brain2024}, and social networks \cite{social2020,social2023}. Graph Signal Processing (GSP) has become widespread, extending classical signal processing techniques rooted in the Fourier Transform to include the Graph Fourier Transform (GFT) \cite{GFT2020}, graph fractional Fourier transform \cite{GFFT2024}, graph signal sampling \cite{Gsamp2021}, and graph aggregated sampling \cite{GAsamp2023}. Time-vertex graph signals (TVGS) naturally extend graph signals by associating each graph vertex with a time-series signal \cite{overview2023}. This structure enables data processing from both vertex and time perspectives. Consequently, TVGS is well-suited for representing and analyzing irregular time-varying data, leading to the development of TVGS processing \cite{2016JFT,timevertex,ji2019}, including graph filtering \cite{filter1,filter2}, data inference \cite{infer1,infer2}, and spatiotemporal graph learning \cite{learn1,learn2}. Applications in domains such as computer vision \cite{CV}, anomalous sound detection \cite{sound}, and oceanography \cite{Qiu} stand to benefit significantly from advances in TVGS processing.

Efficient sampling and reconstruction of TVGS are crucial for accurately capturing and analyzing real-world phenomena. Since the topology of TVGS handles data from both the vertex and temporal perspectives simultaneously, the volume of data grows exponentially \cite{timevertex,ji2019}. This presents significant challenges in data acquisition, storage, and processing. Therefore, sampling theory is a critical aspect of time-point graph signal processing, aiming to recover the entire original signal from partial observations. This paper focuses on the sampling and reconstruction problem of finite time-vertex graph signals (FTVGS). In FTVGS, each vertex is associated with a finite-length time sequence, allowing the signal to be modeled as a matrix \cite{sheng2024sampling}.

Discrete signal processing (DSP) and GSP have been combined to introduce the joint time-vertex Fourier transform (JFT), which provides the spectral expansion of FTVGS \cite{2016JFT}. Based on the known JFT spectrum, some progress has been made in sampling and reconstructing FTVGS \cite{FTVGSsamp2018,sheng2024sampling,FTVGSsamp2023,FTVGSsamp2024}. We call these methods deterministic sampling. However, in practice, we often cannot obtain the spectral information of the signal before sampling. For example, in sensor networks, each row of a matrix represents the time series collected by a sensor, while each column represents a traditional graph signal. Before deploying the sensors, the JFT spectrum of the FTVGS is unknown, making sampling methods that rely on known JFT spectra impractical.

In this case, we consider using random sampling to address the challenge of unknown JFT spectra. One approach is to sample the entire FTVGS through an unbiased stochastic process, followed by formulating an optimization problem for reconstruction based on assumptions such as low-rank or smoothness \cite{TVGSre2017,TVGSre2019,TVGSre2023}. However, this method assumes that all elements of the FTVGS are sampled with the same probability, which is often unrealistic due to factors such as participant willingness or environmental conditions. Lorenzo et al. designed a probabilistic sampling strategy where each vertex at every time instant is sampled with a given probability \cite{adaptive}. Cai et al. proposed the cross-concentrated sampling (CCS) strategy to ensure certain elements remain unobserved \cite{CCS}. However, these sampling schemes do not reduce the number of nodes and time instants generating samples.

In practice, we prefer to sample only a subset of rows and columns, while others produce no samples at all due to hardware, environmental, or cost limitations. We propose a \textit{subset random sampling scheme} for FTVGS, significantly reducing the number of rows and columns producing samples. That is, we randomly select a subset of rows and columns from the matrix to form a submatrix and then randomly sample within the submatrix.

The subset random sampling scheme is straightforward but results in samples with entire rows and columns missing, which we call structural missing. In \cite{sheng2024subset}, we first proposed the subset random sampling scheme and established theoretical guarantees for reconstructing the original FTVGS under structural missing. This paper provides more detailed theoretical analyses and proofs regarding the sufficient conditions for guaranteed reconstruction with high probability. 

However, most existing algorithms for FTVGS reconstruction from randomly observed samples rely on the uniform sampling assumption. Some studies reconstruct the original signal by relaxing the rank minimization problem \cite{SVT,reweighted,NCMC}. Qiu \textit{et al}. extended the definition of smooth signals from static graph to FTVGS, utilizing the smoothness of temporal differences for FTVGS reconstruction \cite{TVGSre2017}. To accelerate algorithm convergence, Giraldo \textit{et al}. proposed a reconstruction method based on the Sobolev smoothness of FTVGS \cite{TRSS}. The analysis and comparison of reconstruction algorithms are detailed in Table~\ref{tab:diff}.

Existing algorithms are rarely designed for structural missing data. To address the reconstruction of signals from samples restricted to the submatrix, we introduce a reconstruction framework with low-rank, sparse, and smooth priors (LSSP), which was not discussed in our previous work \cite{sheng2024subset}. To our knowledge, no existing reconstruction algorithm has leveraged all three signal priors simultaneously. The algorithm within the LSSP framework efficiently recovers the original FTVGS even when rows and columns are missing. 

In brief, the contributions of this paper are summarized as follows:
\begin{itemize}
    \item We propose a more practical subset random sampling scheme for FTVGS. In this scheme, samples are only observed from the submatrix of FTVGS, effectively reducing the number of sampled vertices (rows) and instants (columns).

    \item We prove sufficient conditions for reconstructing the original FTVGS from samples obtained through subset random sampling schemes, including lower bounds on the number of selected rows and columns (\cref{lem:IJ}), as well as a lower bound on the number of observed samples (\cref{thm_S}).

    \item We introduce an LSSP reconstruction framework that deeply exploits the prior information of the FTVGS to address the challenges posed by structural missing. The feasibility of the reconstruction and the effectiveness of the framework are verified through experiments.
\end{itemize}

The rest of the paper is organized as follows. In \cref{sec:rela}, we introduce the theoretical and algorithmic studies on random sampling and reconstruction of FTVGS. In \cref{sec:model} we introduce the FTVGS model. In \cref{sec:sample}, we present the subset random sampling scheme for FTVGS and discuss the conditions for reconstructing the original FTVGS. In \cref{sec:recon}, we design an LSSP reconstruction framework tailored to the samples with structural missing. Numerical results are provided in \cref{sec:experiment}, and conclusions are drawn in \cref{sec:conclusion}.

The notation in this paper is as follows. Flowery letters are used to denote sets. 
For a matrix $\bX \in \Complex^{N \times T}$, let $\bX(i, j)$ denotes the $(i, j)$-th entry of $\bX$, and $\bX(\calI, :)$, $\bX(:, \calJ)$, and $\bX(\calI, \calJ)$ denote the row submatrix with row indices $\calI$, the column submatrix with column indices $\calJ$, and the submatrix of $\bX$ composed of rows indexed by $\calI$ and columns indexed by $\calJ$, respectively.
$\sigma_i(\bX)$ represents the $i$-th largest singular value of $\bX$, while $\lambda_i(\bX)$ is the $i$-th smallest eigenvalue of $\bX$. $|| \bX ||_F = \left( \sum_i \sum_j |\bX(i,j)|^2 \right)^{1/2}$ denotes the Frobenius norm.

\section{Related Works}
\label{sec:rela}

Sampling methods that utilize prior information about the signal to determine specific sample indices are known as deterministic sampling. In contrast, methods that do not predefine the sampling locations are referred to as random sampling. This section provides a brief review of research related to both sampling approaches, with the main focus of this paper being the exploration of random sampling theory for FTVGS.

\subsection{Deterministic sampling of FTVGS}

Based on the known JFT spectral support, the separate sampling scheme applies graph sampling in the vertex and time domains of FTVGS, respectively, resulting in a submatrix containing a part of rows and columns as samples \cite{FTVGSsamp2018}. However, there remains the potential to reduce the number of samples further. Sheng \textit{et al}. jointly processed the FTVGS spectra in both vertex and time domains, established the lower bound of the total sampling density based on the JFT spectrum, and proposed a feasible multi-band sampling scheme \cite{sheng2024sampling}.

In addition, Wei \textit{et al}. designed a reconstruction scheme for FTVGS using subspace prior knowledge and proposed a generalized sampling framework for FTVGS by applying the sampling framework of shift-invariant spaces \cite{FTVGSsamp2024}. Xiao \textit{et al}. proposed a sampling method for FTVGS based on known graph spectrum support and introduced a robust reconstruction algorithm leveraging the smoothness of the signal \cite{FTVGSsamp2023}.

The above-mentioned studies are sampling and reconstruction of FTVGS based on the spectrum of signals. However, in practice, the spectral information is unknown prior to sampling. Therefore, we consider random sampling of FTVGS when the JFT spectrum is unavailable.

\subsection{Random Sampling of FTVGS}

Random sampling of FTVGS is similar to the problem of random sampling of matrices. The pioneering work \cite{MC1} explored the matrix completion (MC) problem based on the Bernoulli sampling model, where the sample index set is obtained through Bernoulli sampling. Moreover, \cite{MC1} and \cite{MC2} proved a lower bound on the number of samples required for accurate recovery with high probability.

Subsequent research on MC expanded beyond Bernoulli sampling to uniform sampling with and without replacement \cite{MC2,noisyMC}. In addition, \cite{MC2010} demonstrated the equivalence of Bernoulli sampling and uniform sampling in MC. Moreover, the lower bound on the number of samples required for signal reconstruction is further reduced in \cite{MC2021}.

\begin{figure}[htbp]
    \begin{center}
        \includegraphics[scale=0.45]{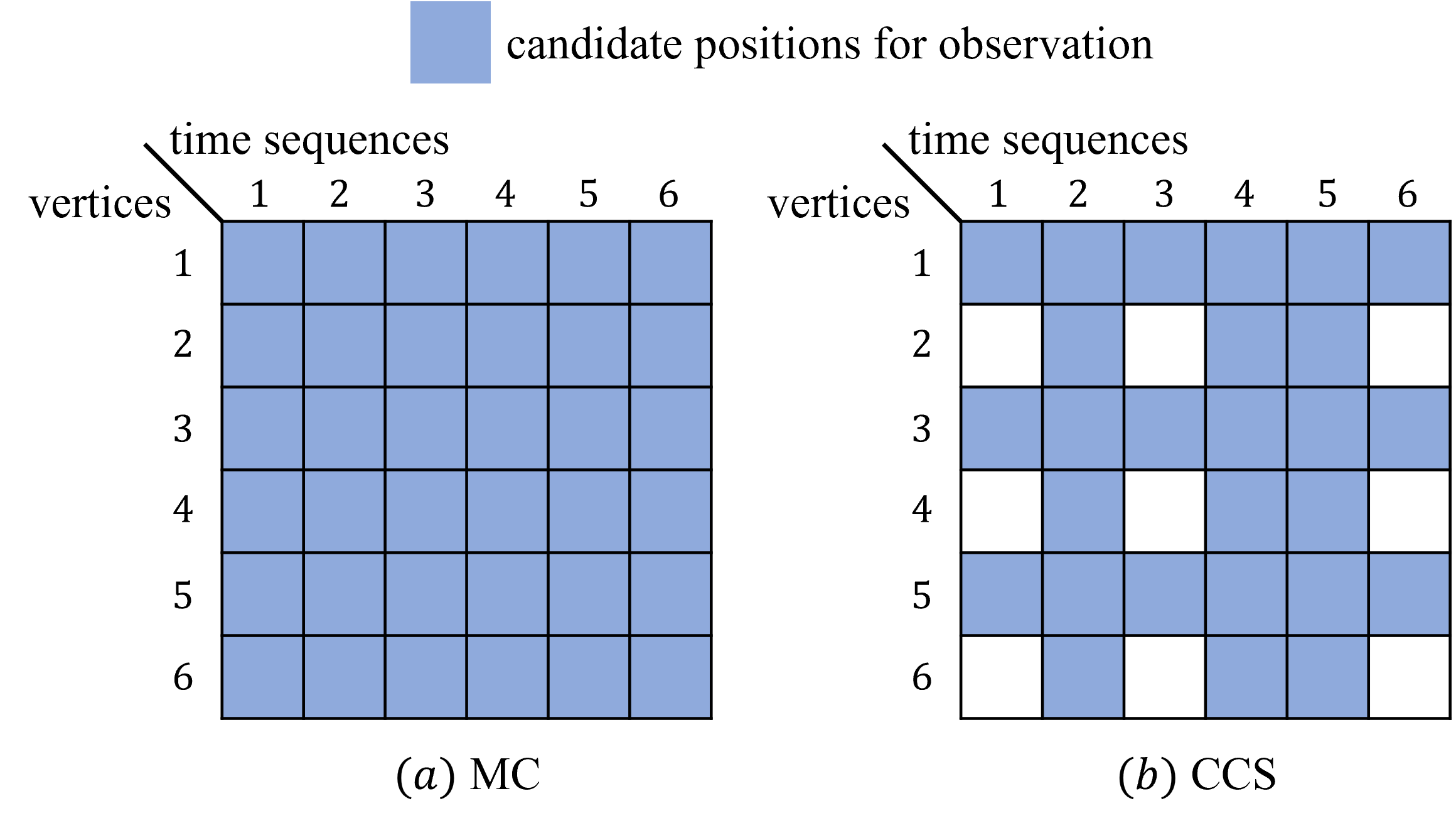}   
    \end{center}
    \caption{The visualization of candidate positions for observation.}
    \label{fig:area}
\end{figure}

Numerous studies on sampling and reconstruction of FTVGS have adopted an approach based on sampling via an unbiased stochastic process, leading to reconstruction algorithms that leverage priors such as smoothness \cite{TVGSre2017,TVGSre2019,TVGSre2023}. Although these studies provide insight into the sampling and reconstruction of FTVGS, they assume that all elements of the matrix are sampled with the same probability (\cref{fig:area} (a)), which is not always feasible in practice. For example, in sensor networks, practitioners typically aim to deploy sensors at a subset of candidate locations while minimizing operational durations. 

A subsequent study combined uniform sampling with CUR decomposition, proposing CCS \cite{CCS}. Specifically, CCS randomly samples within row and column submatrices of FTVGS to obtain the samples (\cref{fig:area} (b)). While CCS ensures that some elements of the FTVGS remain unobserved, it still requires that every vertex and instant produce samples.

In contrast to existing methods, our proposed subset random sampling scheme restricts the sample indices to a subset of rows and columns of FTVGS, ensuring that no samples are produced from the unselected rows and columns (\cref{fig:sub_samp} ). This is motivated by practical scenarios where sampling is often desired only at certain vertices and instances. For example, in sensor networks, we aim to deploy as few sensors as possible. Besides, hardware limitations may prevent sampling at every instance.

To address structurally missing samples in subset random sampling, we introduce an LSSP reconstruction framework and validate its feasibility and effectiveness. The LSSP framework integrates low-rank property, smoothness, and the correlations between rows and columns of FTVGS.

\section{Model}
\label{sec:model}

An undirected graph can be represented as $\calG = (\calV_\calG, \calE_\calG)$, where $\calV_\calG$ is the set of vertices with $|\calV_\calG| = N$, $\calE_\calG=\{ e_1, \dots, e_M \}$ is the set of edges. We consider every vertex in $\calV_\calG$ relates to a time sequence of length $T$, which can be represented by a directed cyclic graph $\calT = (\calV_\calT, \calE_\calT)$. Such signals are FTVGS \cite{timevertex}, whose graphical model is $ \calG \times \calT$ (as shown in \cref{fig:sub_samp}), where $\times$ is the Cartesian product. An FTVGS is given to the matrix $\bX \in \Complex^{N \times T}$. Each row of $\bX$ is a finite time sequence on a vertex, and each column is a static graph signal. 

Suppose each edge of $\calG$ is assigned an orientation, which is arbitrary but fixed. We can define the incidence matrix $\bQ \in \Real^{N \times M}$ as
\begin{equation*}
    \bQ (i,m) = \left \{ 
        \begin{array}{ll}
        1,  & {\text{if $i \in \calV_\calG$ is the initial vertex of edge $e_m$};} \\ 
        -1, & {\text{if $i$ is the terminal vertex of edge $e_m$};} \\
        0,  & {\text{otherwise}.} 
        \end{array}\right.
\end{equation*}
Then the gradient on $\calG$ can be defined as $\bQ^T \bX$ and the graph Laplacian matrix is $\bL_\calG = \bQ \bQ^T$ \cite{incidence}. Let $l \sim i$ indicate that there is an edge between vertices $l, i \in \calV_\calG$. A typical measure of how much a signal varies on $\calG$ for every $j \in \calV_\calT$ is the quadratic
$$ \bX(:, j)^T \bL \bX(:, j) = \frac{1}{2} \sum_{i \in \calV_\calG} \sum_{l \sim i} (\bX(l, j)-\bX(i, j))^2,$$
which is the sum of the neighborhood variation of all vertices, i.e., graph total variation (TV).

Then we introduce first and second-order temporal differential operators
\begin{equation*}
    \bD_1=\left[\begin{matrix}
    -1 &      &     &     \\
    1  &   -1 &     &     \\
       &   1  &   \ddots  &     \\
       &      &   \ddots &  -1 \\
       &      &          &  1
    \end{matrix} \right],
    \bD_2=\left[\begin{matrix}
    1  &      &     &     \\
    -2 &   1  &     &     \\
    1  &   -2 &     &     \\
       &   1  &   \ddots &   \\
       &      &   \ddots &  -2\\
       &      &   \ddots &  1
    \end{matrix} \right],
\end{equation*}
where $\bD_1 \in \Real^{T \times (T-1)}$ and $\bD_2 \in \Real^{T \times (T-2)}$. For each $i \in \calV_\calG$, the first and second order differences of the signal on $\calT$ can be represented as \cite{TV2014}
$$(\bX \bD_1)(i,j) = -\bX(i,j)+\bX(i,j+1), $$ 
and
$$(\bX \bD_2)(i,j) =\bX(i,j)-2\bX(i,j+1)+\bX(i,j+2). $$
Subsequently, the gradient of $\bX$ in the graph and time domains is
\begin{equation*}
    \triangledown \bX(i,j) = \left[\begin{matrix}
    (\bQ^T \bX)(i,j) \\
    (\bX \bD_1)(i,j)
    \end{matrix} \right].
\end{equation*}

Our objective is to propose a subset random sampling scheme for FTVGS and to prove the lower bound on the number of samples required for random sampling on the submatrix of FTVGS. So we begin with a formal expression of the widely used assumptions.  

\begin{assumption}
\label{as_r}
     \cite{MC1,XRC2X} (low-rank, smoothness) An FTVGS $\bX \in \Complex^{N \times T}$ is a rank-$r$ matrix, where $r< \min \{ N,T \}$. For all $ i \in \calV_\calG, j \in \calV_\calT$, 
    $$\max \left\{ \left| \bX(:,j)^T \bL \bX(:,j) \right|, || \triangledown \bX(i,j) ||, \left| (\bX \bD_2)(i,j) \right| \right\} \leq C $$
    for some constant $C$.
\end{assumption}

The low rank implies that $\bX$ can be expressed as a matrix with only $r$ non-zero rows (or columns) through a row (or column) linear transformation. This is consistent with the assumption of bandlimited JFT spectra in the sampling and reconstruction theory of FTVGS, as the JFT is a type of linear transformation \cite{sheng2024sampling}. In particular, the JFT spectrum can be expressed as $\Psi_\calG^T \bX \Psi_\calT$, where $\Psi_\calG$ and $\Psi_\calT$ are the GFT basis matrix and discrete Fourier transform (DFT) basis matrix, respectively. The JFT spectral bandlimitness then implies that each row and column of $\Psi_\calG^T \bX \Psi_\calT$ is tending to be sparse.

The quadratic $\bX(:,j)^T \bL \bX(:,j)$ reflects how the $\bX(:,j)$ varies with respect to the underlying graph structure, while the time-domain difference $(\bX \bD_2)(i,j)$ captures the changes in the $\bX(i,:)$ over time. The gradient $\triangledown \bX(i,j)$ indicates both the difference between an element and its neighboring vertices and the variation across time. The fact that quadratic on the graph $\calG$, second-order difference on $\calT$, and gradient are all bounded implies that $\bX$ is smooth. That is, there are no abrupt changes in $\bX$ in either graph or temporal domains. Correspondingly, in the context of TVGS processing theory, the smoothness of $\bX$ somehow indicates that its JFT spectrum exhibits low-frequency characteristics.

Since $\bX$ is rank-$r$, its thin singular value decomposition (SVD) is
$$ \bX = \bU \bSigma \bV^T, $$
where $\bU \in \Real^{N \times r}$, $\bSigma \in \Real^{r \times r}$ and $\bV \in \Real^{T \times r}$. Then we have the incoherence assumption.

\begin{assumption}
\label{as_mu}
    \cite{MC1,MC2} (incoherence) An FTVGS $\bX \in \Complex^{N \times T}$ satisfying \cref{as_r} is $\{ \mu_1, \mu_2 \}$-incoherence for some constants $1 \le \mu_1(\bX) \le \frac{N}{r}, 1 \le \mu_2(\bX) \le \frac{T}{r}$ provided
    $$\frac{N}{r} || \bU ||_{2,\infty}^2 \le \mu_1(\bX), \frac{T}{r} || \bV ||_{2,\infty}^2 \le \mu_2(\bX), $$
    where $|| \bU ||_{2,\infty} := \max_i \left(\sum_j (\bU(i,j))^2 \right)^{1/2}$ denotes the largest row-wise $\ell_2$-norm..
\end{assumption}

Coherence indicates whether the energy of $\bX$ is uniformly distributed across columns and rows. Low coherence means that the structure of $\bX$ is relatively uniform, making it easier to obtain uniform samples during random sampling.

\section{Subset Random Sampling}
\label{sec:sample}

\subsection{Sampling Scheme}

\begin{figure*}[hbp]
    \begin{center}
        \includegraphics[scale=0.55]{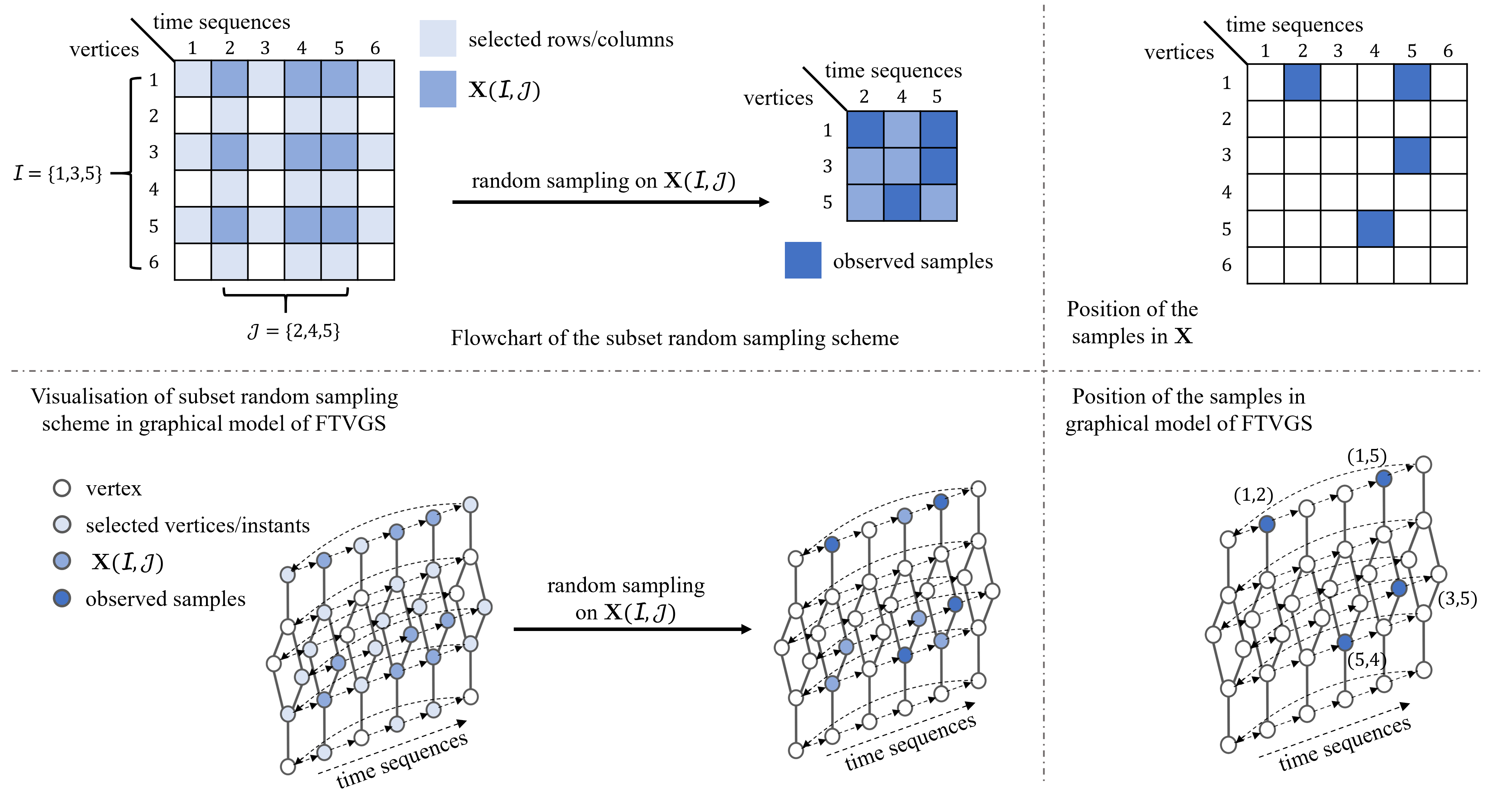}   
    \end{center}
    \caption{Illustration of subset random sampling. }
\label{fig:sub_samp} 
\end{figure*}

As mentioned in the sensor network example in \cref{sec:introduction}, not all sensors can reliably collect data at every moment. In practice, it is preferable to activate only a subset of sensors and sample at a portion of the moments. To address this need, we propose \textit{a subset random sampling scheme} for signals, as illustrated in \cref{fig:sub_samp}. 

Formally, we first randomly select a subset of rows and columns from the FTVGS without replacement, \textit{i.e.}, 
$$\calI \subseteq \calV_\calG, \calJ \subseteq \calV_\calT.$$
Let $\bX(\calI, \calJ)$ be the submatrix of FTVGS $\bX$ composed of rows indexed by $\calI$ and columns indexed by $\calJ$, the samples observed from $\bX$ be indexed by a set $\calS$, and $|\calS|$ denote the number of samples. Subsequently, we uniformly observe samples within the submatrix $\bX(\calI, \calJ)$ with replacement (\cref{fig:sub_samp}), \textit{i.e.},
$$ \calS = \{ (i, j): i \in \calI, j \in \calJ \}. $$
Denote $\bX(i, j), (i, j) \in \calS$ as $\bX_\calS$.

\begin{algorithm}[htbp] 
\caption{Subset random sampling}
\label{proc:samp}
\begin{algorithmic}[1]

\REQUIRE $\bX$

\STATE Uniformly random sample row indices $\calI$ without replacement, yielding $\bX_R = \bX(\calI, :)$.

\STATE Uniformly random sample column indices $\calJ$ without replacement, getting $\bX_{RC} = \bX(\calI, \calJ)$.

\STATE Uniformly random sample entries in $\bX_{RC}$ and get $\calS$.

\ENSURE $\calI$, $\calJ$, $\calS$, $\bX_\calS$

\end{algorithmic}
\end{algorithm}

Unlike existing methods, we would like to sample randomly on the submatrix $\bX(\calI, \calJ)$ instead of $\bX$, ensuring that no samples are observed on unselected rows and columns. We summarize the sampling process as \cref{proc:samp}.

The subset random sampling scheme is straightforward but violates the common assumption in MC problems that each row and column must produce at least one sample. To the best of our knowledge, theoretical guarantees for reconstructing the original FTVGS when the samples are structurally missing in entire rows and columns have not been explored. Therefore, we then study subset random sampling from a theoretical perspective.

\subsection{Sufficient Conditions}

The conditions for FTVGS reconstruction consist of two parts: the lower bounds on the number of retained rows and columns, and the lower bound on the total number of samples observed. First, consider the lower bounds for $|\calI|$ and $|\calJ|$. If too few rows or columns are retained, even when $|\calI| < r$ or $|\calJ| < r$, it becomes challenging to ensure accurate FTVGS reconstruction, even if we observe the entire $\bX(\calI, \calJ)$.

The following lemma ensures that the rank of $\bX(\calI, \calJ)$ remains $r$ with high probability, which is crucial for the subsequent theoretical derivation. \cref{lem:IJ} also provides the lower bounds on the number of selected rows and columns. The proof of \cref{lem:IJ} is presented in APPENDIX~\ref{proof:lem_IJ}.

\begin{lemma}
\label{lem:IJ}
    For an FTVGS $\bX \in \Complex^{N \times T}$ satisfying \cref{as_r} and \cref{as_mu}, let $\bX_R$ and $\bX_{RC}$ be obtained by \cref{proc:samp}. Then if
    $$ |\calI| \ge 3r\mu_1(\bX) \frac{\ln{(2r/\delta)}}{\epsilon^2}, |\calJ| \ge 3r\mu_2(\bX) \frac{\ln{(2r/\delta)}}{\epsilon^2} $$
    for some $0 < \delta < 1$ and $0 < \epsilon < 1$, $\mathrm{rank}(\bX_R) = r$ holds with probability at least $1-\delta$ and $\mathrm{rank}(\bX_{RC}) = r$ holds with probability at least $(1-\delta)^2$.
\end{lemma}

With \cref{lem:IJ}, we obtain the submatrix $\bX_{RC}$. Following \cref{proc:samp}, we will subsequently sample randomly on $\bX_{RC}$. We wish to prove a lower bound on the number of samples required, which depends not only on the rank of $\bX_{RC}$ but also on its incoherence properties. To this end, we prove \cref{lem:mu}, which demonstrates how the properties of the original signal $\bX$ are converted to those of its submatrix $\bX_{RC}$. 

Let $\bX_R = \bU(\calI, :) \Sigma \bV^T$. Then $\bU(\calI, :) \Sigma$ has the thin SVD $\bU_R \Sigma_R \tilde{\bV}^T_R$, where $\tilde{\bV}_R \in \Real^{r \times r}$ is an orthonormal matrix and $\bU_R \in \Real^{r \times |\calI|}$. The thin SVD of $\bX_R$ is 
\begin{equation}
\label{eq:SVD_XR}
    \bX_R = \bU(\calI, :) \Sigma \bV^T = \bU_R \Sigma_R \tilde{\bV}^T_R \bV^T = \bU_R \Sigma_R \bV^T_R,
\end{equation}
where $\bV_R = \bV \tilde{\bV}_R $ is a matrix with orthogonal columns. Likewise, we have $\bX_{RC} = \bU_R \Sigma_R \bV_R(\calJ, :)^T$ and the thin SVD $\Sigma_R \bV_R(\calJ, :)^T = \tilde{\bU}_{RC} \Sigma_{RC} \bV_{RC}^T$, where $\tilde{\bU}_{RC} \in \Real^{r \times r}$ is an orthonormal matrix and $\bV_{RC} \in \Real^{r \times |\calJ|}$. The thin SVD of $\bX_{RC}$ is 
\begin{equation}
\begin{aligned}
    \label{eq:XRC_SVD}
    \bX_{RC} &= \bU_R \Sigma_R (\bV_R(\calJ, :))^T = \bU_R \tilde{\bU}_{RC} \Sigma_{RC} \bV_{RC}^T \\ &= \bU_{RC} \Sigma_{RC} \bV_{RC}^T,
\end{aligned}
\end{equation}
where $\bU_{RC} = \bU_R \tilde{\bU}_{RC} $ is a matrix with orthogonal columns.

\begin{lemma}
\label{lem:mu}
    Under the assumption of \cref{lem:IJ}, if $\calI$ and $\calJ$ satisfy the inequality of \cref{lem:IJ}, then the following conditions hold with probability at least $(1-p)^2, p = r \left[ \frac{e^{-\eta}}{(1-\eta)^{1-\eta}} \right]^{\log r}$:
    $$ || \bU_{RC} ||_{2,\infty} \le \kappa(\bX) \sqrt{\frac{\mu_1(\bX) r}{(1-\eta) |\calI|}}, $$
    $$ || \bV_{RC} ||_{2,\infty} \le \frac{\kappa(\bX)}{1-\eta} \sqrt{\frac{\mu_2(\bX) Nr}{|\calI| |\calJ|}}, $$
    where $\kappa(\bX)$ is the condition number of $\bX$ and $\eta \in [0, 1)$. 
\end{lemma}

The proof of \cref{lem:mu} is presented in APPENDIX ~\ref{proof:lem_mu}.
\cref{lem:mu} shows how the properties of the original signal $\bX$ are converted to those of its submatrix $\bX(\calI, \calJ)$. 

Having \cref{lem:IJ} and \cref{lem:mu}, we then introduce the second part of the conditions for FTVGS reconstruction: the lower bound on the number of samples $|\calS|$. This forms the main theorem of this paper, following the theory of MC problems. But here we observe $\calS$ from $\bX_{RC}$, rather than from $\bX$. 

\begin{theorem}
\label{thm_S}
    For an FTVGS $\bX \in \Complex^{N \times T}$ satisfying \cref{as_r} and \cref{as_mu}, suppose that $\bX_{RC}$ and $\calS$ is derived from \cref{proc:samp}. Then if 
    \begin{align*}
        |\calS| \ge \frac{32 \beta (\kappa(\bX))^4 r^2 N}{(1-\eta)^3} \mu_1(\bX) \mu_2(\bX) \frac{|\calI|+|\calJ|}{|\calI|} \log^2(2n),
    \end{align*}
    where $\beta > 1$, $\kappa(\bX)$ is the condition number of $\bX$, $\eta \in [0, 1)$ and $n = \max{\{|\calI|,|\calJ|\}}$, the $\bX$ can be recovered from $\calS$ with probability at least 
    $$ (1-\delta)^2 (1-p)^2 -\frac{6 \log{n}}{(|\calI|+|\calJ|)^{2\beta-2}} -n^{2-2\beta^{1/2}}. $$
\end{theorem}

The proof of \cref{thm_S} is presented in APPENDIX \ref{proof:thm_S}.

\cref{lem:IJ} and \cref{thm_S} provide the conditions for reconstructing the original FTVGS from the subset random sampled samples. Among the probabilities in \cref{thm_S}, $(1-\delta)^2$ is the probability that guarantees $\mathrm{rank}(\bX_{RC}) = \mathrm{rank}(\bX) = r$ and $(1-p)^2$ is the probability that guarantees the incoherence of $\bX_{RC}$.

We summarize the differences between random sampling of MC, CCS, and subset random sampling in \cref{tab:diff}. It can be observed that when $N$ and $T$ are sufficiently large, uniform random sampling requires a relatively smaller number of samples, while the sample complexities of CCS and subset random sampling are comparable. It is because uniform sampling imposes no constraints, assuming each vertex at each instant is sampled with equal probability, whereas CCS and subset random sampling restrict observable positions. \textit{It is worth noting that subset random sampling does not have a higher sample complexity than CCS, although it constrains the vertices and instants that produce samples to a subset. }
For smaller $N$ and $T$, the parameters of subset random sampling may increase the required number of samples, resulting in a higher lower bound. This is attributed to the fact that increased sampling constraints necessitate more samples to uniquely determine the original signal.

\begin{table*}[htbp]
    \centering
    \caption{Differences between three schemes for random sampling of FTVGS.}
    \begin{tabular}{l|c|c|c}
        \toprule
         &  Random sampling of MC    & CCS    & Subset random sampling \\
        \midrule
        Assumption & \multicolumn{2}{c|}{rank-$r$, $\{ \mu_1, \mu_2 \}$-incoherence}   & rank-$r$, smooth, $\{ \mu_1, \mu_2 \}$-incoherence \\
        \midrule
        \makecell[l]{Sufficient \\ sample complexity} & \makecell[c]{$O(r^2 n_2 \log(n_2) )$ \cite{MC2021}\\ ($n_2 = \max{\{N, T \}}$) } & $O(r^2 n_2 \log^2(n_2) )$ \cite{CCS} & $O(r^2 N \log^2(n))$  \\
        \bottomrule
    \end{tabular}
    \label{tab:diff}
\end{table*}

Our proposed subset random sampling scheme restricts the sample indices within $(\calI, \calJ)$, meaning that no samples will be drawn from unselected rows and columns. However, this comes at the cost of some reconstruction error, as detailed in \cref{sec:experiment}. Because there are rows and columns without samples, MC methods and CCS cannot complete the reconstruction. Methods based on TV inpainting and our proposed reconstruction framework can recover the FTVGS, but cannot completely eliminate the error.

\section{Reconstruction Framework}
\label{sec:recon}

In \cref{sec:sample}, we established the conditions for reconstructing the original FTVGS from samples indexed by $\calS$ through theoretical analysis. In this section, together with \cref{sec:experiment}, we aim to verify the feasibility of FTVGS reconstruction from the perspective of the reconstruction algorithm.

MC has been a topic of extensive research, and well-established frameworks now exist. To address the challenge of reconstructing a complete signal from a limited number of randomly observed samples, an effective approach is to incorporate priors or regularization. A common prior assumes that the original matrix is low-rank or approximately low-rank. By modeling the MC process as a low-rank matrix approximation, it can then be solved using optimization methods.

Our \cref{proc:samp} meets the practical need to reduce the number of vertices and moments to be sampled, but it introduces the issue of structural missing, where entire rows and columns are absent (see \cref{fig:sub_samp}). Reconstructing the FTVGS from these structurally incomplete samples goes beyond the scope of traditional MC conditions and algorithms. This is because the low-rank prior and its variants are insufficient in regularizing structural incompleteness, necessitating additional regularization to compensate for the limitations of low-rankness in the presence of structural missing.

To tackle the challenges posed by structural missing, we aim to more fully exploit the prior information of the FTVGS. Based on our assumptions about the FTVGS, we derive three types of regularization:
\begin{enumerate}
    \item $f_r(\bX)$ stands for rank regularization of the FTVGS. Based on the general \cref{as_r}, we assume that the original signal exhibits high redundancy and can be represented in a more compact form, implying that FTVGS is either low-rank or approximately low-rank. To some degree, low-rank characterizes the global prior of the FTVGS, so the underlying matrix is regularized by the low-rank prior.

    \item $f_c(\bF_\calG, \bF_\calT)$ denotes a priori regularization based on the spatial and temporal correlations of FTVGS. The low-rank property of FTVGS indicates strong correlations between its rows and columns, allowing $\bX$ to be represented in a low-dimensional form through linear transformations. Based on TVGS processing theory, we apply the GFT and DFT bases to obtain a low-dimensional representation of $\bX$. In the graph domain, we have $\bX = \Psi_\calG \bF_\calG$, where $\bF_\calG$ is the GFT spectrum, and each $\bF_\calG(:,j)$ is the bandlimited spectrum of the graph signal $\bX(:,j)$. In the time domain, $\bX^T = \Psi_\calT \bF_\calT$, where each $\bF_\calT(:,i)$ represents the bandlimited spectrum of the time sequence $\bX(i,:)$. To incorporate the sparse prior of FTVGS, we introduce sparse regularization of the graph frequency and the time frequency.

    \item $f_s(\bX)$ represents smoothing regularization. In \cref{as_r}, we also introduced the smooth feature of the FTVGS. This implies that, in both the graph and time domains, the difference between an entry and its neighboring entries is typically small. Commonly studied regularization includes TV and its variants, as well as the smoothness of temporal difference signals.
\end{enumerate}

Now, the method based on three kinds of regularization can be modeled as the following optimization problem
\begin{equation}
    \begin{aligned}
        &\min_{\hat{\bX}} \quad f_{\text{obj}} = f_r(\hat{\bX}) + \gamma_c f_c(\bF_\calG, \bF_\calT) + \gamma_s f_s(\hat{\bX})  \\
        & \begin{array}{c@{\quad \quad \quad}l@{\quad}l}
        s.t. & \hat{\bX} = \Psi_\calG \bF_\calG, \\
             & \hat{\bX}^T = \Psi_\calT \bF_\calT, \\
             & \hat{\bX}(i,j) = \bX(i, j), (i, j) \in \calS,
        \end{array}
    \end{aligned}
\label{eq:optm_0}
\end{equation}
where $\gamma_c$ and $\gamma_s$ are non-negative regularization parameters. 

Some existing studies have not investigated the theoretical guarantees of \cref{proc:samp}, but many reconstruction algorithms incorporating multiple priors have been proposed. In the LSSP reconstruction framework, we summarize some of these existing methods in \cref{tab:sum_alg}.

\begin{table*}[htbp]
    \centering
    \caption{Existing FTVGS reconstruction algorithms.}
    \begin{tabular}{llll}
        \toprule
         & \multicolumn{1}{c}{$f_r(\hat{\bX})$}    &  \multicolumn{1}{c}{$f_c(\bF_\calG, \bF_\calT)$}   & \multicolumn{1}{c}{$f_s(\hat{\bX})$} \\
        \midrule
        SVT \cite{SVT} & nuclear norm $\parallel \hat{\bX} \parallel_* = \sum_i \sigma_i(\hat{\bX}) $  &  \multicolumn{1}{c}{-} & \multicolumn{1}{c}{-} \\
        \midrule
        NC-MC \cite{NCMC} & \makecell[l]{non-convex regularizer $g(\lambda_i (\hat{\bX}^T \hat{\bX}))$, \\ $g(x) = \log(x^{1/2}+1)$} & \multicolumn{1}{c}{-} & \multicolumn{1}{c}{-} \\
        \midrule
        ReLaSP \cite{ReLaSP} & nuclear norm $\parallel \hat{\bX} \parallel_* $ & \makecell[l]{$|| \bF_\calG||_1 = \max_j \sum_i^N |\bF_\calG(i,j)|$ \\ or $\parallel \bF_\calT \parallel_1 $} & \multicolumn{1}{c}{-} \\
        \midrule
        LIMC \cite{TV2018} & nuclear norm $\parallel \hat{\bX} \parallel_* $ & \multicolumn{1}{c}{-} & \makecell[l]{$||\hat{\bX} \bD_2||_F^2 + ||\hat{\bX}^T \bD_2'||_F^2$ \\ $\bD_2'$ is the second-order graph differential operator} \\
        \midrule
        LRDS \cite{TVGSre2019} & nuclear norm $\parallel \hat{\bX} \parallel_* $ & \multicolumn{1}{c}{-} & $\text{tr}((\hat{\bX} \bD_1)^T \bL \hat{\bX} \bD_1)$ \\
        \midrule
        LRGTS \cite{TVGSre2023} & nuclear norm $\parallel \hat{\bX} \parallel_* $ & \multicolumn{1}{c}{-} & $||\hat{\bX} \bD_2||_F^2 + \text{tr}(\hat{\bX}^T \bL \hat{\bX} )$ \\
        \bottomrule
    \end{tabular}
\label{tab:sum_alg}
\end{table*}

In Table~\ref{tab:sum_alg}, the low-rank variant serves as a global regularization by constraining the low-rank structure of the underlying matrix \cite{SVT,NCMC}. Building on the low-rank prior, ReLaSP introduced methods that impose $\ell_1$ regularization on the graph domain spectrum, encouraging each $\bF_\calG(:,j)$ to be sparse \cite{ReLaSP}. Although they only constrained a single domain, this approach can easily be extended to the graph and time domains framework, where $f_c(\bF_\calG, \bF_\calT) = || \bF_\calG||_1 + ||\bF_\calT||_1$. In this case, each $\bF_\calG(:,j)$ and each $\bF_\calT(:,i)$ tend to be sparse.

On the other hand, some studies introduced a smoothness prior in addition to the low-rank prior. LIMC constrains the smoothness constraints of FTVGS in both the time and graph domains using a modified second-order TV \cite{TV2018}. Similarly, LRGTS aims to leverage second-order smoothness information of FTVGS in both domains, denoting the smoothing in the graph domain by $\text{tr}(\hat{\bX}^T \bL \hat{\bX} )$ \cite{TVGSre2023}. LRDS defines the smoothness of the temporal differences of FTVGS as $\text{tr}((\hat{\bX} \bD_1)^T \bL \hat{\bX} \bD_1)$ and uses it as the smoothing regularization \cite{TVGSre2019}.

Since no existing research has designed optimization problems that combine $f_r(\hat{\bX})$ with both $f_c(\bF_\calG, \bF_\calT)$ and $f_s(\hat{\bX})$, we present an optimization problem that simultaneously incorporates all three regularizations for a more comprehensive comparison and analysis. Combined with the reweighting method  \cite{reweighted,reweighted_l1}, we express the regularization function of the optimization problem as
\begin{equation}
    \begin{aligned}
        f_{\text{obj}} & = g(\lambda_i (\hat{\bX}^T \hat{\bX})) + \gamma_\calG || \bW_\calG \odot \bF_\calG||_1 \\
        & \quad + \gamma_\calT || \bW_\calT \odot \bF_\calT||_1 + \gamma_d ||\hat{\bX} \bD_2||_F^2, \\
    \end{aligned}
\label{eq:optm}
\end{equation}
where $g(x) = \log(x^{1/2}+1)$, $\odot$ denotes the element-wise multiplication of two matrices, $\bW_\calG$ and $\bW_\calT$ are adjustable weight matrices for sparsity terms. 

In considering smoothing regularization, it is found that in practice, FTVGS tends to exhibit stronger smoothness in the time domain and less smoothness in the graph domain \cite{TVGSre2017}. Therefore, we adopt a temporal smoothing prior. \textit{It is noteworthy that Eq.~\ref{eq:optm} does not entirely disregard the low graph-domain smoothness.} The term $|| \bF_\calG||_1 $ makes the graph signal band-limited at each instance, implying inherent correlations among signals at vertices. This implicitly constrains graph-domain smoothness in some sense. The solution to the optimization problem based on \cref{eq:optm} is detailed in APPENDIX \ref{sec:algo}, and the resulting algorithm is summarized in \cref{algo_LSSP}.

\begin{algorithm}[htbp] 
\caption{LSSP}
\label{algo_LSSP}
\begin{algorithmic}[1]

\REQUIRE observed data $\bX_\calS$, bases matrices $\Psi_\calG, \Psi_\calT$, temporal differential operator $\bD_2$, regularization parameters $\gamma_\calG, \gamma_\calT, \gamma_d$, $\zeta > 0$, $\mu_1^{(1)} = \mu_2^{(1)} = \mu_3^{(1)} > 0$, $\rho_1 = \rho_2 = \rho_3 > 1$

\STATE \textbf{Initialize}: initialize $\hat{\bX}^{\{1\}}, \bF_\calG^{\{1\}}, \bF_\calT^{\{1\}}, \bE^{\{1\}}, \bY_1^{(1)}, \bY_2^{(1)}, \bY_3^{(1)}$ as matrices filled with ones, $p = 1$, $\bW_\calG^{\{1\}}(i, j) = \bW_\calT^{\{1\}}(i, j) = 1$

\WHILE{not converged}

\FOR{$k = 1, 2, \dots, K$}
  \STATE $b_1^1 = 1, \bF_\calG^{(k),1} = \bF_\calG^{\{p\},(k)}, \bA_1^1 = \bF_\calG^{\{p\},(k)}$
  \FOR{$q = 1, 2, \dots, Q$}
  \STATE update $\bF_\calG^{(k),q+1}$ by \cref{eq:FGkq+1,eq:FGkA}
  \ENDFOR
  \STATE $\bF_\calG^{\{p\},(k+1)} = \bF_\calG^{(k),Q}$

  \STATE $b_2^1 = 1, \bF_\calT^{(k),1} = \bF_\calT^{\{p\},(k)}, \bA_2^1 = \bF_\calT^{\{p\},(k)}$
  \FOR{$q = 1, 2, \dots, Q$}
  \STATE update $\bF_\calT^{(k),q+1}$ by \cref{eq:FTkq+1}
  \ENDFOR
  \STATE $\bF_\calT^{\{p\},(k+1)} = \bF_\calT^{(k),Q}$

  \STATE update $\hat{\bX}^{\{p\},(k+1)}$ by \cref{eq:hatXk+1}
  \STATE update $\bE^{\{p\},(k+1)}$ by \cref{eq:Ek+1}
  \STATE update $\bY_1^{(k+1)}, \bY_2^{(k+1)}, \bY_3^{(k+1)}, \mu_1^{(k+1)}, \mu_2^{(k+1)}, \mu_2^{(k+1)}$ by \cref{eq:Ymuk+1}
  
\ENDFOR

\STATE $\hat{\bX}^{\{p+1\}} = \hat{\bX}^{\{p\},(K)}, \bF_\calG^{\{p+1\}} = \bF_\calG^{\{p\},(K)}, \bF_\calT^{\{p+1\}} = \bF_\calT^{\{p\},(K)}, \bE^{\{p+1\}} = \bE^{\{p\},(K)}$

\STATE update $\bW_\calG^{\{p+1\}}, \bW_\calT^{\{p+1\}}$ by \cref{eq:Wp+1}
\STATE $p = p+1$

\ENDWHILE

\ENSURE reconstructed FTVGS $\hat{\bX}$

\end{algorithmic}
\end{algorithm}

Overall, since traditional MC algorithms cannot handle the problem of structurally missing samples, we delve deeper into the prior information of FTVGS. Starting from basic assumptions, we classify the priors into three types and organize them into an LSSP reconstruction framework. Within this framework, we first introduce some existing reconstruction algorithms and then present \cref{eq:optm}, which incorporates all three kinds of regularizers.

\section{Experiment}
\label{sec:experiment}

In this section, we verify the feasibility of reconstructing the FTVGS from samples indexed by $\calS$ by applying different reconstruction methods to real-world datasets. Furthermore, we analyze different algorithms within the LSSP framework based on the experimental results.

In the sampling phase, we follow the subset random sampling scheme (\cref{proc:samp}), which ensures that no sample is observed from $\bX(\calV_\calG \backslash \calI, \calV_\calT \backslash \calJ)$. For each $\bX$, we sample at different ratios, where $\alpha_{RC}$ represents the sampling ratio of randomly selected rows or columns, \textit{i.e.}, 
$$|\calI| = \frac{{\rm Round}(\alpha_{RC} N)}{N}, |\calJ| = \frac{{\rm Round}(\alpha_{RC} T)}{T},$$ where ${\rm Round}(\cdot)$ represents rounding numbers to the nearest integer. $\alpha_{\bX_{RC}}$ represents the ratio of uniform sampling on $\bX_{RC}$, \textit{i.e.}, 
$|\calS| = {\rm Round}(\alpha_{\bX_{RC}} |\calI| |\calJ|) / ( |\calI| |\calJ|)$. And $\alpha_{total} = |\calS|/(NT)$ is the total sampling ratio.

For reconstruction in all experiments, the parameters of \cref{algo_LSSP} were set as follows: $\zeta = 0.001, \rho_1 = \rho_2 = \rho_3 = 0.01$, $\mu_1^{(1)} = \mu_2^{(1)} = \mu_3^{(1)} = \frac{1}{2 \max(\sigma(\bX_\calS))}$, where $\sigma(\bX_\calS)$ is the singular values of $\bX_\calS$.

\subsection{Competing Algorithms and Performance Metrics}

We compare the \cref{algo_LSSP} with five algorithms: ICURC \cite{CCS}, MC-NC \cite{NCMC}, ReLaSP \cite{ReLaSP}, LIMC \cite{TV2018}, and LRDS \cite{TVGSre2019}.

The ICURC algorithm integrates error minimization, low-rank constraints, and CUR decomposition. In a recent study on random sampling theory, CCS \cite{CCS}, ICURC demonstrated strong performance in reconstructing the original signal. Thus, we include it as a comparison algorithm.
    
The four algorithms, MC-NC, ReLaSP, LIMC, and LRDS, listed in \cref{tab:sum_alg}, all leverage partial prior information of the signals to formulate the optimization problem. Optimization problem based on \cref{algo_LSSP} is constructed within the LSSP framework, incorporating all three types of regularities.

To compare the reconstruction effects of different algorithms, we define the normalized root mean square error (NRMSE) to quantify the error between signals $\bX_a$ and $\bX_b$,
$$ {\rm NRMSE}(\bX_a, \bX_b) = \frac{||\bX_a - \bX_b||_F}{||\bX_a||_F}. $$

\subsection{Test on Synthetic Data}

\begin{figure}[htbp]
    \centering
    \subfigure[The energy of generated $\bF_\calJ$]
    {
	    \includegraphics[scale=0.38]{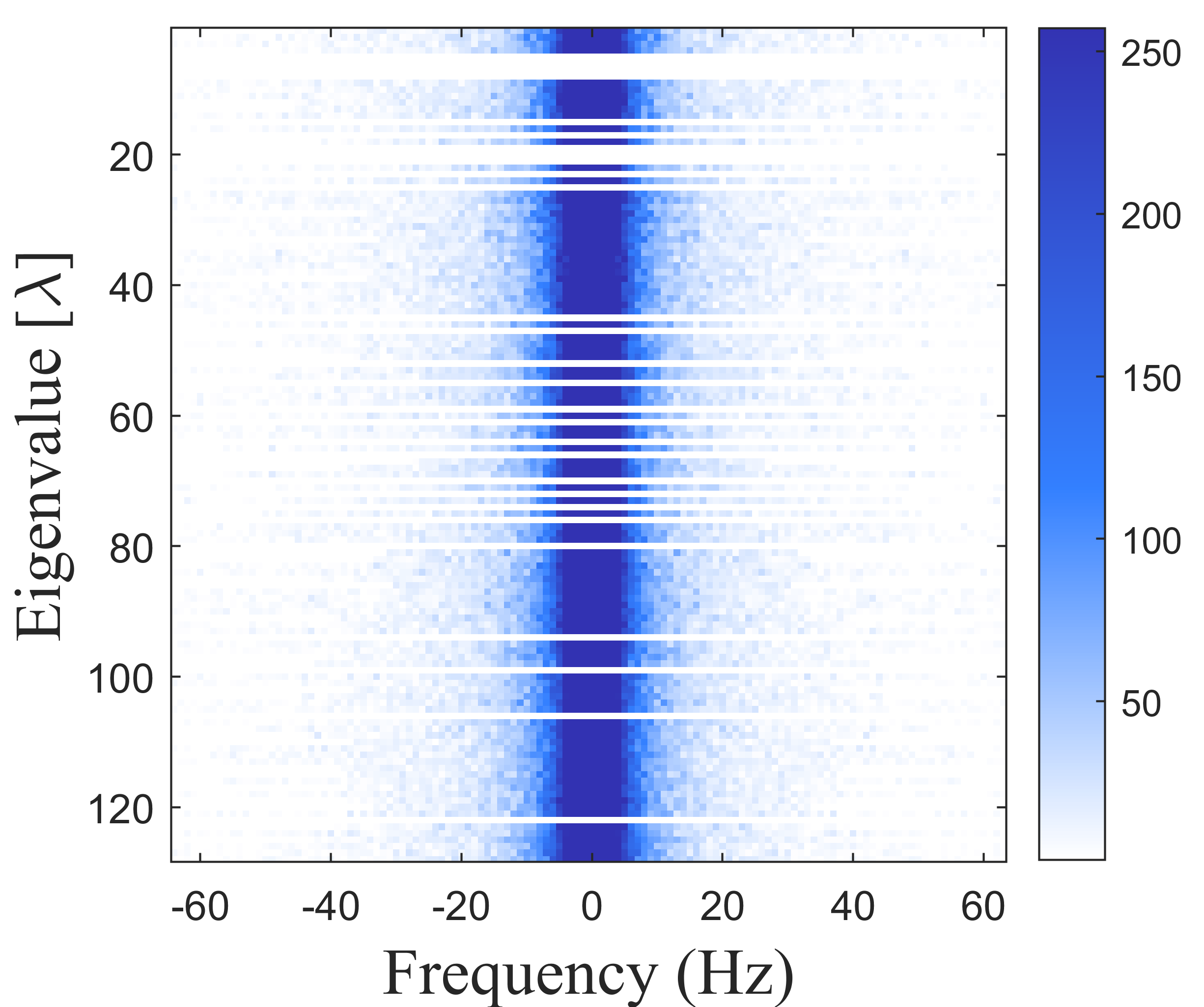}   
	}\\
 
    \subfigure[The energy of $\bF_\calT$]
    {
	    \includegraphics[scale=0.38]{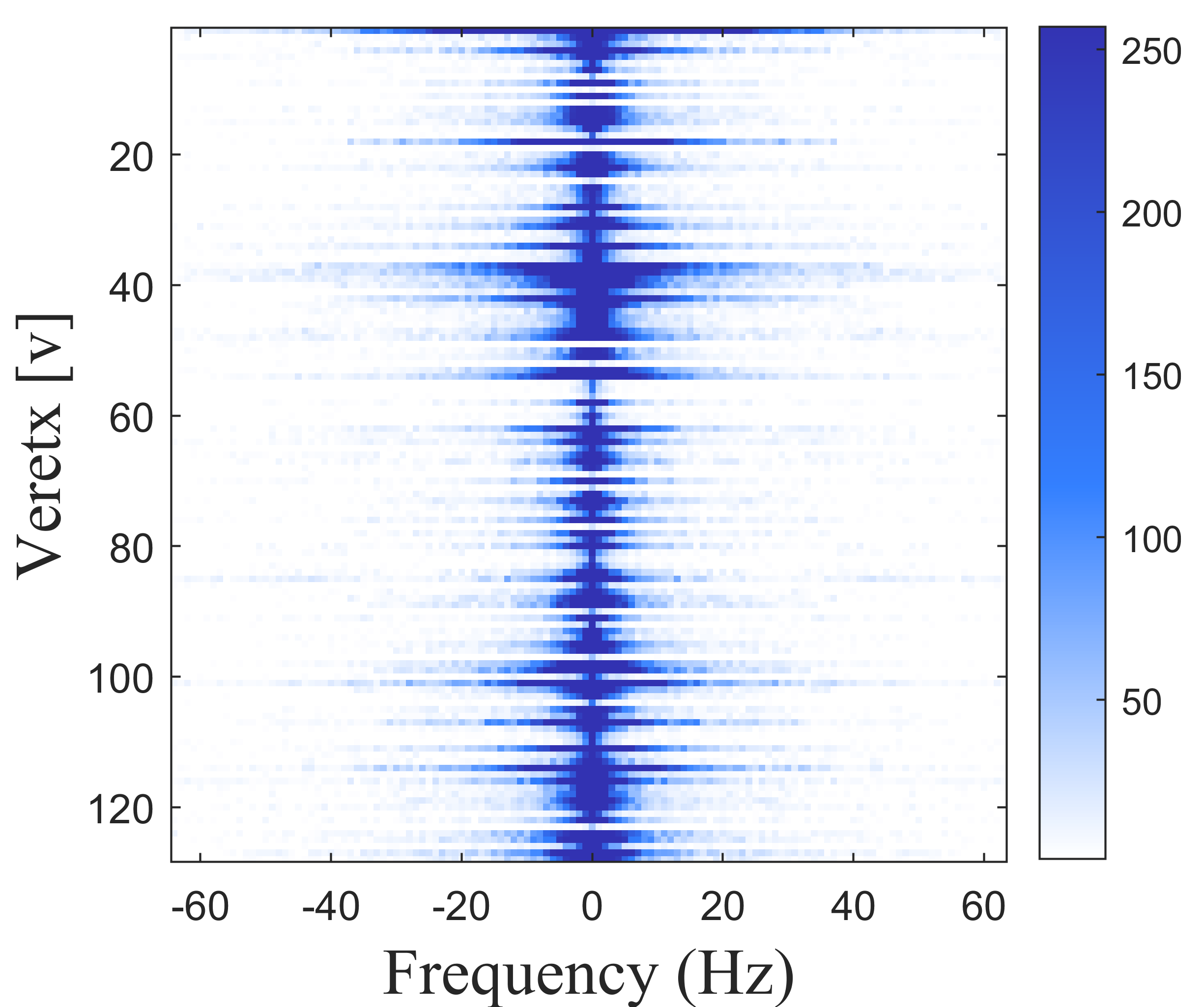}   
    }
    \subfigure[The energy of $\bF_\calG$]
    {
	    \includegraphics[scale=0.38]{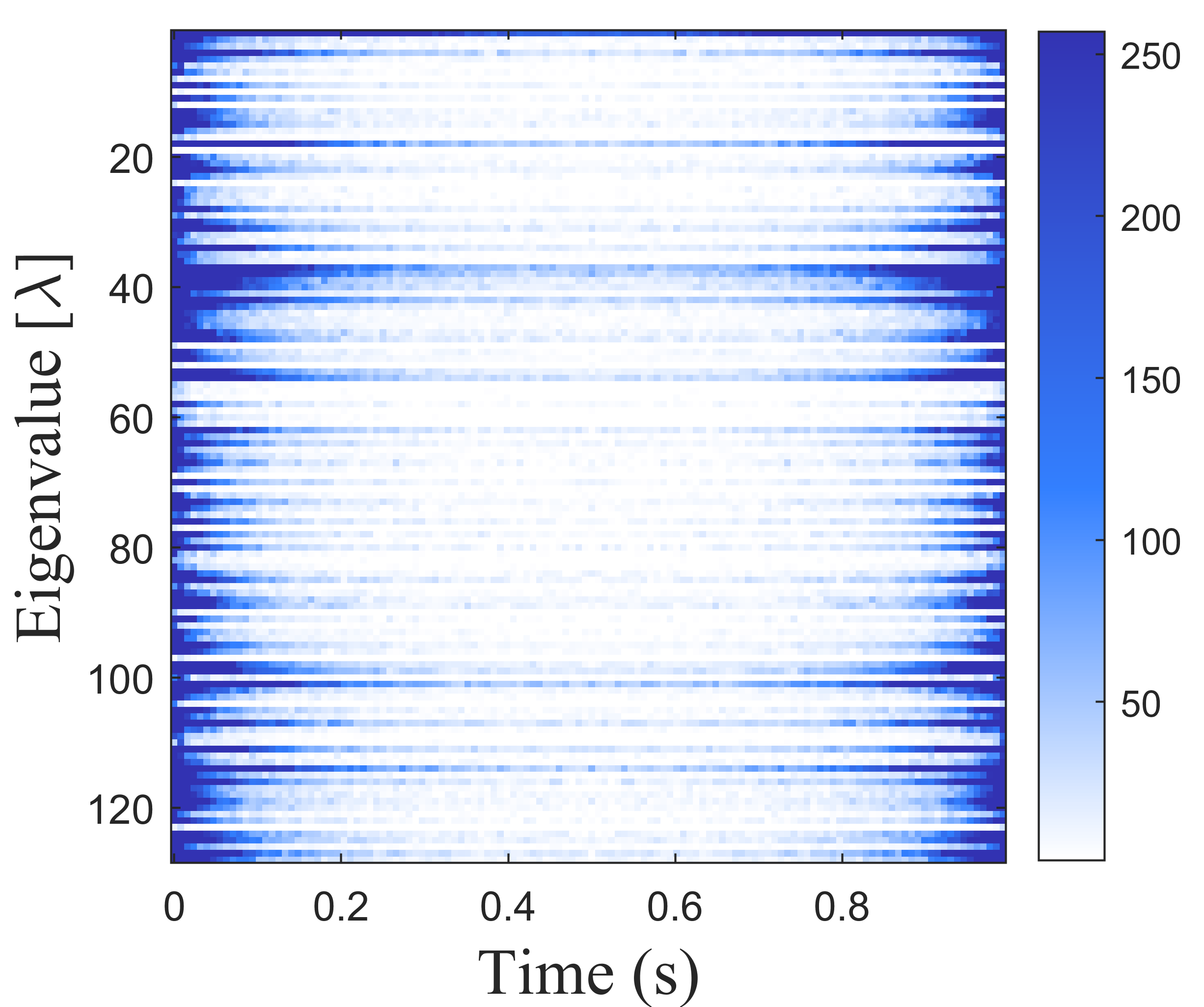}   
    }
    \caption{Visualization of the spectrum of the generated data. }
\label{fig:F_syn} 
\end{figure}

We generate a $128 \times 128$ FTVGS $\bX$ by $\bX = \Psi_\calG \bF_\calJ \Psi_\calT^T$, where $\Psi_\calG$ and $\Psi_\calT$ are given GFT basis and FT basis, respectively. As shown in Fig.~\ref{fig:F_syn} (a), $\bF_\calJ$ represents the generated JFT spectrum with 100 nonzero rows, \textit{i.e.}, 100 nonzero low-frequency spectra. Each non-zero spectral component (each row) in $\bF_\calJ$ is randomly generated, with the low-frequency bandwidth randomly selected between 28 and 88. Subsequently, Fig.~\ref{fig:F_syn} (b) displays $\bF_\calT = \Psi_\calG \bF_\calJ$, demonstrating that the spectrum of each temporal signal $\bX(i,:)$ is bandlimited. In Fig.~\ref{fig:F_syn} (c), $\bF_\calG = \bF_\calJ \Psi_\calT^T$ is also constrained to 100 non-zero rows, indicating that the graph spectrum of each graph signal $\bX(:,j)$ is sparse. All visualizations represent the squared absolute values of the elements.

We generate a total of 100 FTVGS. Under different sampling ratio settings, we calculate the mean NRMSE for various reconstruction algorithms, as compared in \cref{tab:nrmse_synthetic}.

\begin{table}[htbp]
    \centering
    \caption{Comparison of NRMSE for reconstruction algorithms on synthetic data.}
    \begin{tabular}{l|cccc}
        \toprule
        $\alpha_{RC}$ &  $90\%$    & $80\%$    & $70\%$   & $60\%$ \\
        $\alpha_{\bX_{RC}}$ & $90\%$    & $80\%$    & $70\%$   & $60\%$ \\
        $\alpha_{total}$ &    $72.65\%$ & $50.80\%$ & $34.61\%$   & $21.71\%$ \\
        \midrule
        ICURC \cite{CCS}     & 0.4378 & 0.6009 & 0.7149 & 0.8022 \\
        MC-NC \cite{NCMC}    & 0.4545 & 0.5922 & 0.7048 & 0.7976 \\
        LRDS \cite{TVGSre2019} & 0.3651 & 0.4889 & 0.5750 & 0.6555 \\
        LIMC \cite{TV2018}   & 0.1529 & 0.2146 & 0.2852 & 0.3049\\
        ReLaSP \cite{ReLaSP} & 0.1306 & 0.1805 & 0.2269 & 0.2506 \\
        LSSP  & \textbf{0.1118} & \textbf{0.1615} & \textbf{0.1939} & \textbf{0.2330} \\
        \bottomrule
    \end{tabular}
    \label{tab:nrmse_synthetic}
\end{table}

\subsection{Test on Real-world Data}
\label{subsec:LA}

The traffic data METR-LA is collected from loop detectors in the highways of Los Angeles County \cite{data_traffic}. We chose data collected by 207 sensors from March 1st, 2012 to June 27th, 2012 for this experiment. Each sensor is regarded as a vertex $v_i \in \calV_\calG$, and we denote the data of every 512 consecutive timesteps as $\bX$ for simulation. The METR-LA is divided into 100 FTVGS in total. That is, each $\bX$ is an FTVGS with $N = 207$ vertices and $T = 512$ timesteps.

\begin{figure}[htbp]
    \centering 
    \subfigure[The energy of $\bF_\calT$ ]
    {
	    \includegraphics[scale=0.45]{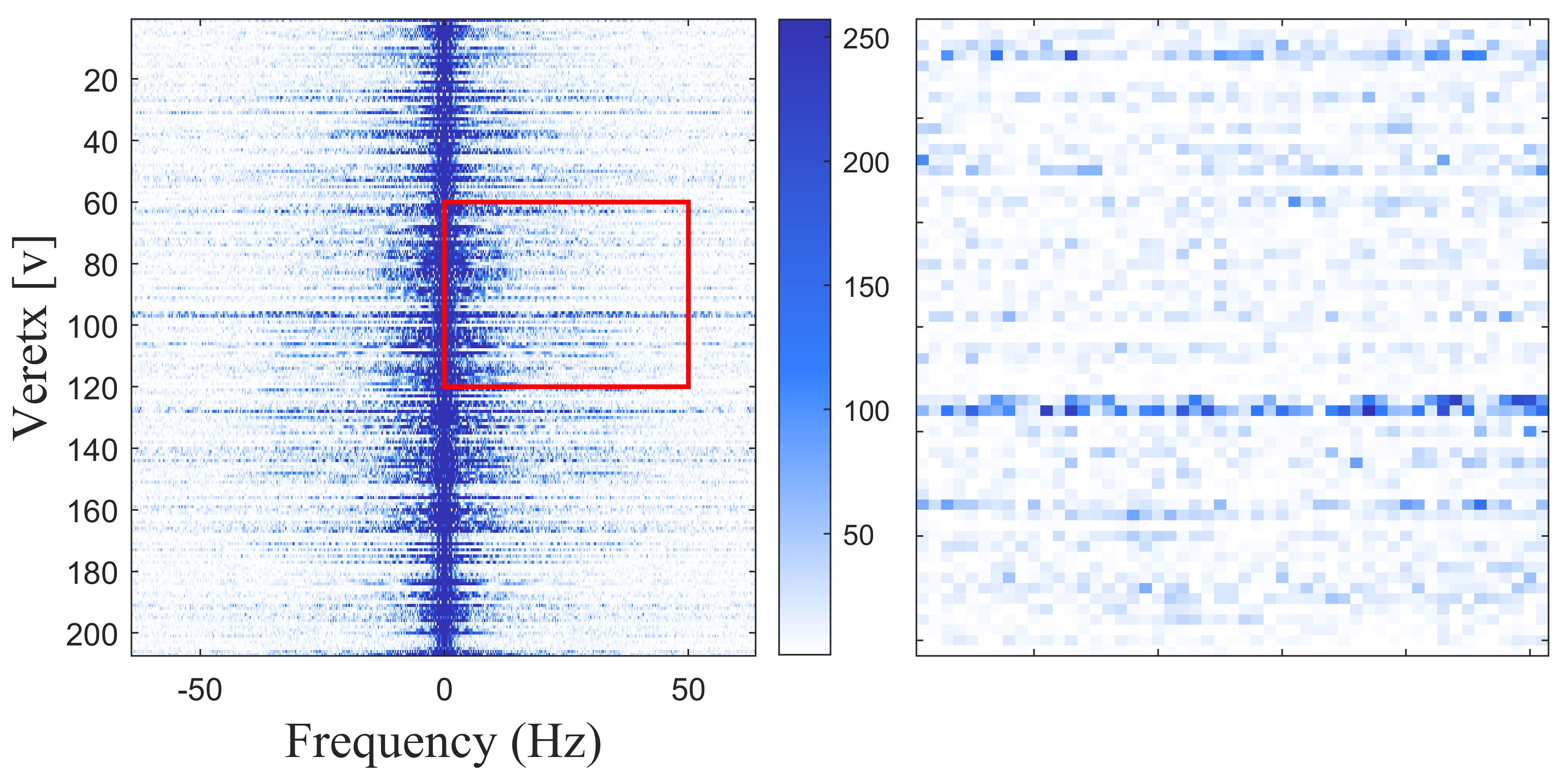}   
    }
    \subfigure[The energy of $\bF_\calG$]
    {
	    \includegraphics[scale=0.45]{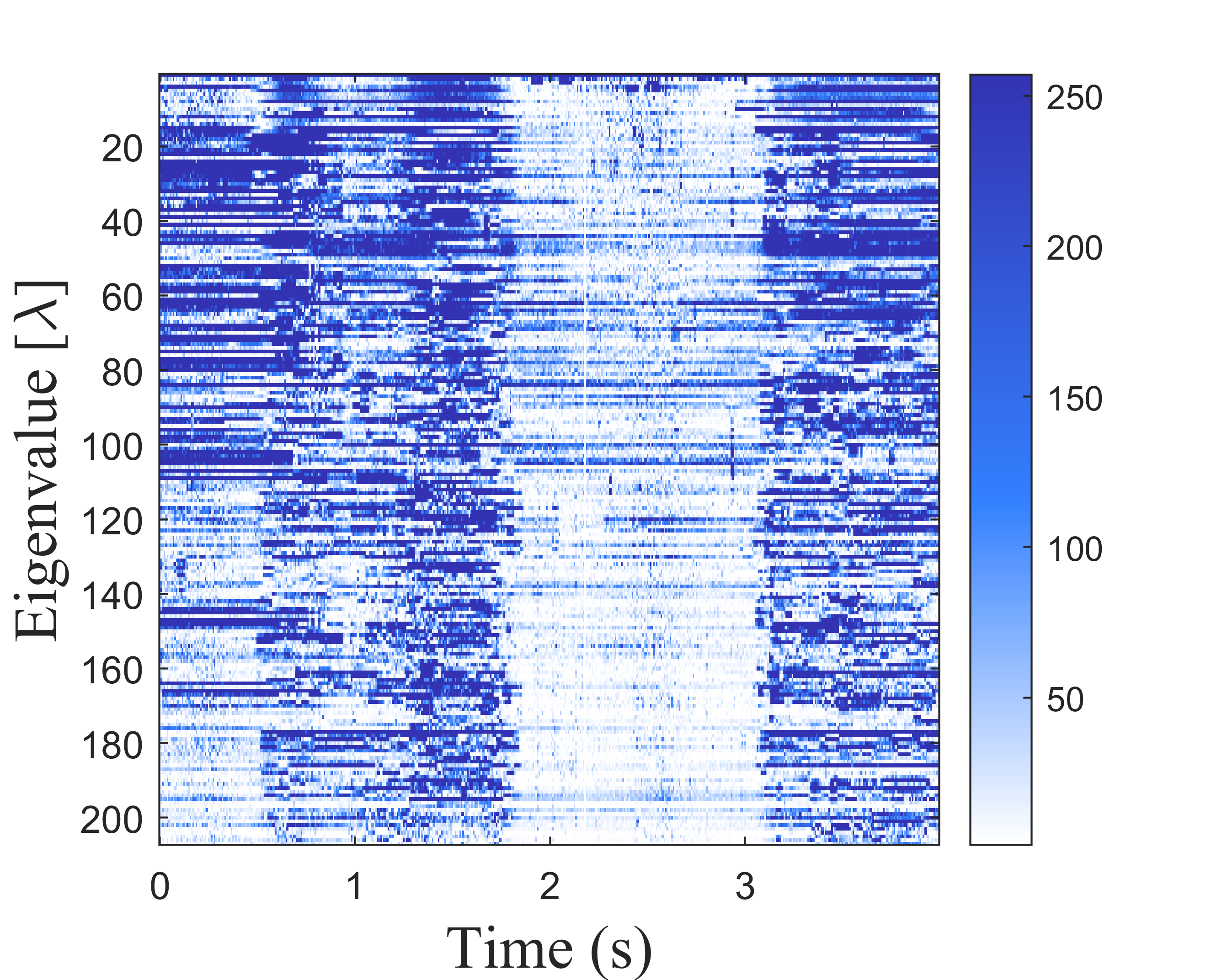}   
    }
    \caption{Visualization of the spectra of $\bX$ getting from the METR-LA data. Partial details in the red rectangle in (a) are enlarged for better visual inspection, displayed in the second column of (a).}
\label{fig:F_LA} 
\end{figure}

For the METR-LA dataset, the energy of $\bF_\calT$ and $\bF_\calG$ are shown in Fig.~\ref{fig:F_LA} (a) and (b), respectively. From Fig.~\ref{fig:F_LA} (a), each sensor acquired signal $\bX(i,:)$ has a spectrum $\bF_\calT(i,:)$ that tends to be low-frequency, which implies both temporal smoothness and spectral sparsity. Furthermore, Fig.~\ref{fig:F_LA} (b) reveals that the graph spectrum of each graph signal $\bX(:,j)$ also exhibits sparsity but does not demonstrate strong smoothness.

Under different sampling ratio settings, we calculate the average NRMSE for various reconstruction algorithms, as compared in \cref{tab:nrmse_LA}

\begin{table}[htbp]
    \centering
    \caption{Comparison of NRMSE for reconstruction algorithms on METR-LA.}
    \begin{tabular}{l|cccc}
        \toprule
        $\alpha_{RC}$ &  $90\%$    & $80\%$    & $70\%$   & $60\%$ \\
        $\alpha_{\bX_{RC}}$ & $90\%$    & $80\%$    & $70\%$   & $60\%$ \\
        $\alpha_{total}$ &    $72.81\%$ & $51.37\%$ & $34.29\%$   & $21.55\%$ \\
        \midrule
        ICURC \cite{CCS}     & 0.4337 & 0.5966 & 0.7110 & 0.8075 \\
        MC-NC \cite{NCMC}    & 0.4375 & 0.5982 & 0.7207 & 0.8063 \\
        LRDS \cite{TVGSre2019} & 0.3931 & 0.5025 & 0.5917 & 0.6669 \\
        LIMC \cite{TV2018}   & 0.1409 & 0.2218 & 0.2739 & 0.3491\\
        ReLaSP \cite{ReLaSP} & 0.1311 & 0.2105 & 0.2296 & 0.2731 \\
        LSSP  & \textbf{0.1248} & \textbf{0.1934} & \textbf{0.2121} & \textbf{0.2457} \\
        \bottomrule
    \end{tabular}
    \label{tab:nrmse_LA}
\end{table}

\begin{figure}[htbp]
    \centering
    \includegraphics[scale=0.6]{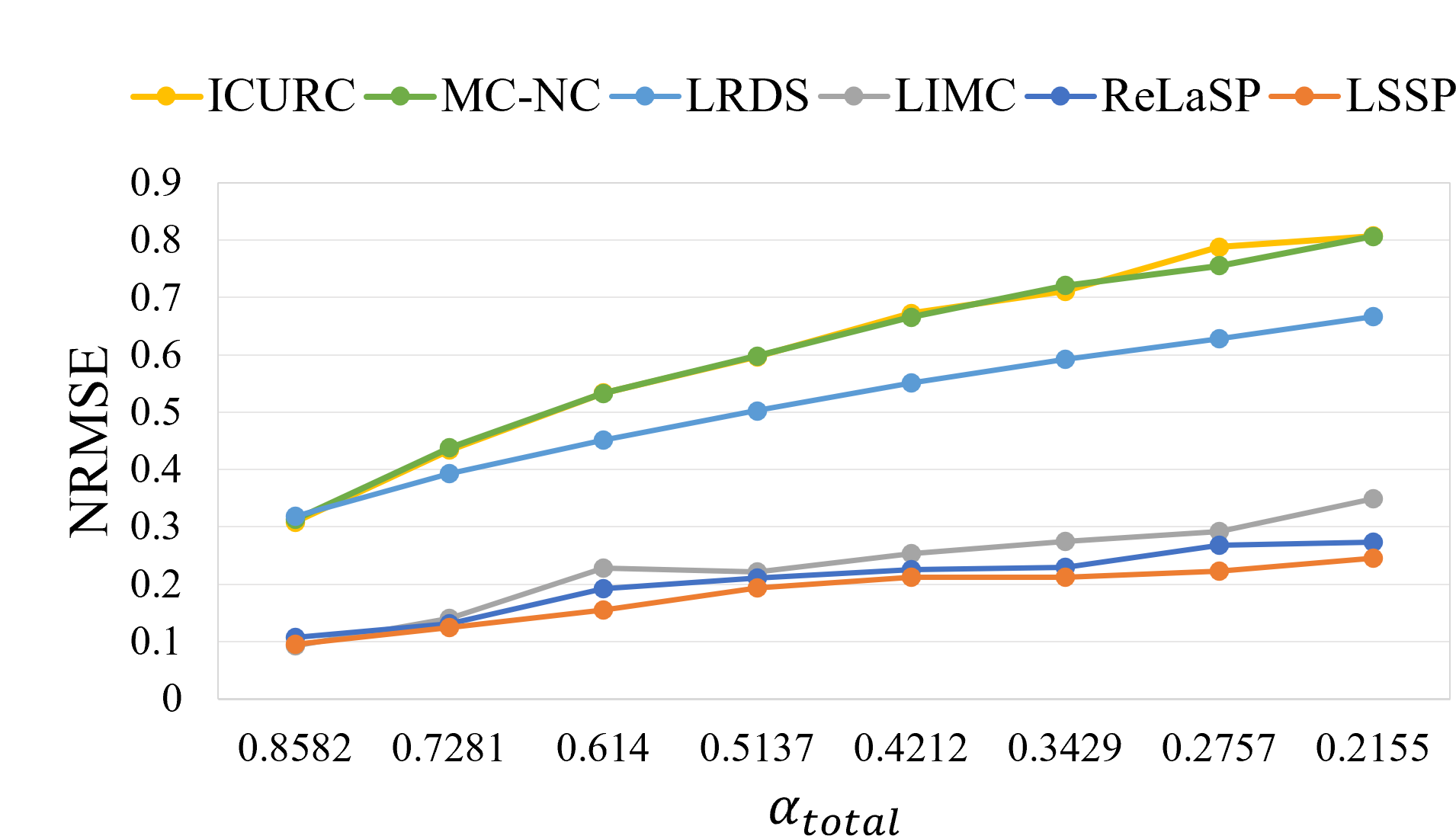}
    \caption{Comparison of NRMSE for reconstruction algorithms on METR-LA.}
    \label{fig:nrmse_LA}
\end{figure}

To investigate the trade-off between sampling rate and reconstruction accuracy, we conducted additional experiments under a wider range of total sampling ratios. The results for different methods are summarized in Fig~\ref{fig:nrmse_LA}.

\begin{figure*}[htbp]
    \centering
    \includegraphics[scale=0.5]{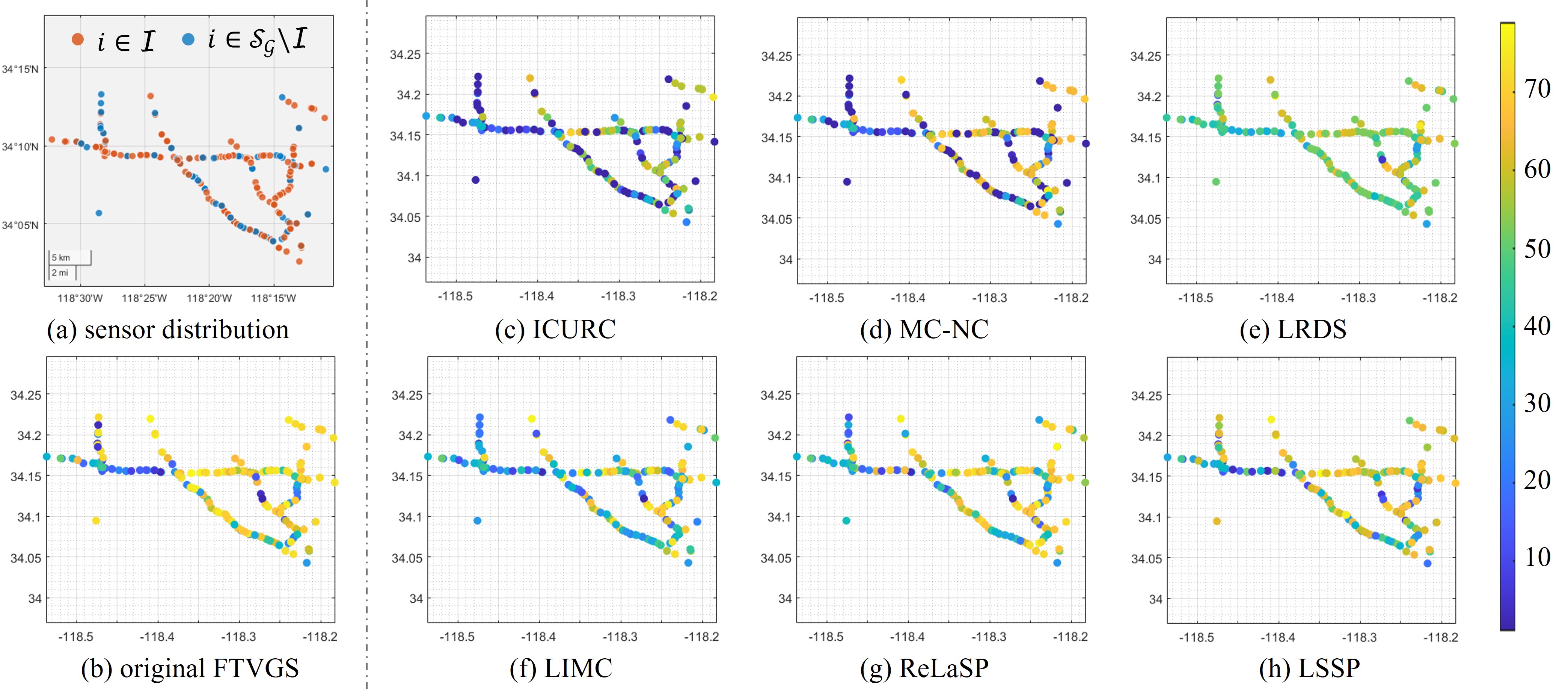}
    \caption{Visual results of signal reconstruction on METR-LA at $j = 100$ when $\alpha_{RC} = 70\%, \alpha_{\bX_{RC}} = 70\%, \alpha_{total} = 34.29\%$. The horizontal and vertical coordinates are longitude and latitude, respectively.}
    \label{fig:visualize}
\end{figure*}

It can be seen that in the task of reconstructing FTVGS from samples with structural missing, LSSP can achieve better results. Focusing on the results of the LSSP, the reconstruction error increases steadily as the sampling rate decreases from $85.82\%$ to $51.37\%$. However, when the total sampling rate further drops from $42.12\%$ to $34.29\%$, the NRMSE remains almost unchanged (0.2119 to 0.2121). This indicates that reducing the sampling rate to around $34.29\%$ does not significantly degrade reconstruction performance, making it a practical choice for costefficient sampling. On the other hand, further reductions below $34.29\%$ result in a sharp increase in error, suggesting that this point serves as a boundary beyond which reconstruction quality deteriorates rapidly. Therefore, we recommend sampling rates of approximately $\alpha_{RC} = 70\%, \alpha_{\bX_{RC}} = 70\%, \alpha_{total} = 34.29\%$ as the optimal balance between sampling cost and reconstruction accuracy. We visualize the comparison between the original FTVGS and the signal reconstructed by different algorithms in Fig.~\ref{fig:visualize}

\subsection{Discussion}

Now, we analyze the experimental results in relation to the FTVGS properties shown in \cref{fig:F_syn,fig:F_LA}.

\begin{enumerate}
    \item $f_r(\bX)$ only: The results in \cref{tab:nrmse_synthetic,tab:nrmse_LA} demonstrate that low-rank matrix approximation models are ineffective in handling data with structural missing. Algorithms like ICURC and MCNC, which rely solely on the low-rank prior of the FTVGS, cannot reconstruct the elements in $\calV_\calG \backslash \calI $ and $\calV_\calT \backslash \calJ $.

    \item $f_r(\bX)+f_s(\bX)$: LRDS and LIMC use two different smoothing priors, respectively. The smoothing regularizer of LRDS does not take into account the time domain, \textit{i.e.}, it fails to reconstruct the elements in $\calV_\calT \backslash \calJ $. LIMC uses the modified second-order TV regularizer, which takes into account the information of the neighboring entries, and thus gives a better result in reconstructing both the missing rows and columns.

    \item $f_r(\bX)+f_c(\bF_\calG, \bF_\calT)$: From \cref{tab:nrmse_synthetic,tab:nrmse_LA}, we can see that each column of $\bF_\calG$ is sparse and each row of $\bF_\calT$ is either low-frequency or tends to be low-frequency, indicating that $\bF_\calT(i,:)$ is also sparse. In the ReLaSP experiments, $\bF_\calG$ and $\bF_\calT$ are used to constrain the correlations between rows and columns, respectively. This allows missing rows and columns to be reconstructed through these correlations.

    \item $f_s(\bX)$ v.s. $f_c(\bF_\calG, \bF_\calT)$: Algorithms like LIMC reconstruct the FTVGS based on a smoothing prior. However, as shown in \cref{fig:F_LA} (b), the FTVGS does not exhibit strong smoothness in the graph domain. Each $\bF_\calG(:,j)$ is only sparse rather than low-rank, meaning its energy is not concentrated solely in the low-frequency (smaller $\lambda$) region. Consequently, LIMC may over-smooth the reconstructed signal in the vertex domain. In contrast, ReLaSP achieves better reconstruction results by imposing a sparse regularity constraint on the graph spectrum $\bF_\calG$.

    \item LSSP: Based on 4), \cref{eq:optm} selects $|| \bF_\calG||_1$ as the regularizer. Additionally, as shown in \cref{fig:F_LA} (a), the FTVGS is low-frequency in the time domain, but real-world signals tend to exhibit sparsity even within low-frequency bands. Therefore, for the regularity of $\bF_\calT$, we incorporate both a smoothing regularizer $||\hat{\bX} \bD_2||_F^2$ and sparsity regularizer $|| \bF_\calT||_1$. Ultimately, \cref{algo_LSSP} achieves better reconstruction results when addressing structurally missing samples. This demonstrates that, within the LSSP framework, leveraging signal priors to refine the optimization problem enhances reconstruction accuracy.
\end{enumerate}

\section{Conclusion}
\label{sec:conclusion}

In this work, motivated by practical applications, we propose a subset random sampling scheme for FTVGS, which involves uniform sampling from the submatrix $\bX_{RC}$ instead of the entire matrix $\bX$. Theoretically, we prove sufficient conditions that guarantee the reconstruction of the original FTVGS with high probability under the subset random sampling scheme. Additionally, for data with randomly missing elements and entire rows and columns, we introduce the targeted reconstruction algorithm LSSP. We validate the effectiveness of the LSSP algorithm on a real-world dataset.

\appendices

\section{Proof of the \cref{lem:IJ}}
\label{proof:lem_IJ}

We first prove that $\mathrm{rank}(\bX_R) = r$ holds with probability at least $1-\delta$ when we sample $|\calI|$ rows of the signal $\bX$ uniformly without replacement. The lower bound of $|\calI|$ is related to $r$ and $\mu_1(\bX)$ as shown in \cref{eq:I}.
From the thin SVD of $\bX$ we know that $\bU \in \Real^{N \times r}$ is a matrix with orthogonal columns. Selecting the rows of $\bX$ is essentially row sampling on $\bU$.
    
    Theorem 4.1 and Corollary 4.2 in \cite{rowsample} proved that $\mathrm{rank}(\bU(\calI, :)) = r$ holds with probability at least $1-\delta$ if
    $$ |\calI| \ge 3N \xi_1 \frac{\ln{(2r/\delta)}}{\epsilon^2} $$
    for some $0 < \delta < 1$ and $0 < \epsilon < 1$, where $\xi_1 = || \bU ||_{2, \infty}^2$.

    According to \cref{as_mu}, $ \mu_1(\bX) r/N \ge || \bU ||_{2, \infty}^2 = \xi_1$, so we have 
    \begin{equation}
        |\calI| \ge 3r\mu_1(\bX) \frac{\ln{(2r/\delta)}}{\epsilon^2}.
        \label{eq:I}
    \end{equation}
    Further, we have $\bX_R = \bU(\calI, :) \Sigma \bV^T$, where $\Sigma \bV^T$ is a matrix with orthogonal rows. Then $\mathrm{rank}(\bX_R) = r$ holds with probability at least $1-\delta$.

    We secondly prove that $\mathrm{rank}(\bX_{RC}) = r$ holds with probability at least $(1-\delta)^2$ when we sample $|\calJ|$ columns of the signal $\bX_R$ uniformly without replacement. The lower bound of $|\calJ|$ is related to $r$ and $\mu_2(\bX)$ as shown in \cref{eq:J}.

    The thin SVD of $\bX_R$ is $\bX_R = \bU_R \Sigma_R \bV^T_R$, as derived in \cref{eq:SVD_XR}. 
    The same argument yields
    $ |\calJ| \ge 3T \xi_2 \ln{(2r/\delta)}/\epsilon^2$
    for some $0 < \delta < 1$ and $0 < \epsilon < 1$, where $\xi_2 = || \bV_R ||_{2, \infty}^2 $ \cite{rowsample}.
    
    We have
    \begin{equation*}
        \xi_2 = || \bV_R ||_{2, \infty}^2 = || \bV \tilde{\bV}_R ||_{2, \infty}^2  \overset{(a)}{=} || \bV ||_{2, \infty}^2 \le \frac{\mu_2(\bX) r}{T},
    \end{equation*}
    where $\tilde{\bV}_R$ is an orthogonal matrix, for any $i$, $||\bV(i,:) \tilde{\bV}_R||_2 = ||\bV(i,:)||_2$ and (a) holds.
    So we get 
    \begin{equation}
        |\calJ| \ge 3r \mu_2(\bX) \frac{\ln{(2r/\delta)}}{\epsilon^2}.
        \label{eq:J}
    \end{equation}

    Since $\bU(\calI, :) \Sigma$ is a matrix with orthogonal columns with probability at least $1-\delta$. Then $\mathrm{rank}(\bX_{RC}) = r$ holds with probability at least $(1-\delta)^2$.

\section{Proof of the \cref{lem:mu}}
\label{proof:lem_mu}

We demonstrate that the submatrix $\bX_{RC}$ exhibits similar incoherence. Using $\bU_R$ and $\bV_R$ as intermediaries, we establish upper bounds on $\bU_{RC}$ and $\bV_{RC}$, illustrating how the properties of the original FTVGS $\bX$ are transferred to its submatrix $\bX_{RC}$.

According to the thin SVD of $\bX_R$, Lemma 4.2 and Lemma 4.3 in \cite{rCUR} give
    \begin{equation}
    \begin{aligned}
    \label{eq:UR1}
        || \bU_R ||_{2, \infty} &\le \kappa(\bX) || \bU(\calI, :)^\dagger ||_2 \sqrt{\frac{\mu_1(\bX)r}{N}} \\
        & = \kappa(\bX) \frac{1}{\sigma_{\mathrm{min}}(\bU(\calI, :))} \sqrt{\frac{\mu_1(\bX)r}{N}}. 
    \end{aligned}
    \end{equation}

    According to Lemma 3.4 in \cite{subsamp_sigma}, 
    \begin{equation}
        \label{eq:UR2}
        \sqrt{\frac{(1-\eta) |\calI|}{N}} \le \sigma_{\mathrm{min}}(\bU(\calI, :))
    \end{equation}
    with failure probability at most $p = r\left[ \frac{e^{-\eta}}{(1-\eta)^{1-\eta}} \right]^{\log r},\eta \in [0,1)$.
    
    Then we substitute \cref{eq:UR2} into \cref{eq:UR1} to get
    \begin{equation*}
    \begin{aligned}
        || \bU_R ||_{2, \infty} &\le \kappa(\bX) \sqrt{\frac{N}{(1-\eta) |\calI|}} \sqrt{\frac{\mu_1(\bX)r}{N}} \\
        & = \kappa(\bX) \sqrt{\frac{1}{1-\eta}} \sqrt{\frac{\mu_1(\bX)r}{|\calI|}}
    \end{aligned}
    \end{equation*}
    holds with probability at least $ 1-p$.
    \cref{eq:XRC_SVD} yields
    \begin{align*}
        || \bU_{RC} ||_{2,\infty} &= || \bU_R \tilde{\bU}_{RC} ||_{2, \infty} \\
        &\overset{(b)}{=} || \bU_R ||_{2, \infty} \le \kappa(\bX) \sqrt{\frac{\mu_1(\bX) r}{(1-\eta) |\calI|}},
    \end{align*}
     where $\tilde{\bU}_{RC}$ is an orthogonal matrix, for any $i$ we have $||\bU_R(i,:) \tilde{\bU}_{RC}||_2 = ||\bU_R(i,:)||_2$ and (b) holds.

    Next, we consider $|| \bV_{RC} ||_{2,\infty}$. According to \cref{eq:XRC_SVD}, we have $\bV_{RC}^T = \Sigma_{RC}^{-1} \bU_{RC}^T \bX_{RC}$. Thus, for any $i$,
    \begin{equation}
    \begin{aligned}
        \label{eq:VRC}
        ||\bV_{RC}^T \be_i||_2 & \le ||\Sigma_{RC}^{-1}||_2 ||\bU_{RC}^T||_2 ||\bX_{RC} \be_i||_2 \\        
        & = \frac{1}{\sigma_{\mathrm{min}}(\bX_{RC})} \sigma_1(\bU_{RC}) ||\bX_{RC} \be_i||_2 \\
        & = \frac{1}{\sigma_{\mathrm{min}}(\bX_{RC})} ||\bX_{RC} \be_i||_2,
    \end{aligned}
    \end{equation}
    where $\be_i$ is a unit vector whose $i$-th element is 1, $\bU_{RC}$ is a column orthogonal matrix and $\sigma_1(\bU_{RC}) = 1$.
    Besides $\bX_{RC} = \bX(:, \calJ) = \bU_R \Sigma_R \bV_R(\calJ,:)^T$, so
    \begin{equation}
        \label{eq:XRC}
        \begin{aligned}
            ||\bX_{RC} \be_i||_2 
            & \le ||\bU_R||_2 ||\Sigma_R||_2 ||\bV_R(\calJ,:)^T \be_i||_2 \\
            &\le \sigma_1(\bU_R) \sigma_1(\bX_R) ||\bV_R(\calJ,:)^T \be_i||_2 \\
            &= \sigma_1(\bX_R) ||\bV_R(\calJ,:)^T \be_i||_2,
        \end{aligned}
    \end{equation}
    where $\bU_R$ is a column orthogonal matrix and $\sigma_1(\bU_R) = 1$.
    Then combining \cref{eq:VRC} and \cref{eq:XRC}, we have
    \begin{equation}
        \begin{aligned}
            \label{eq:VRC2}
            || \bV_{RC} ||_{2,\infty} &\le \frac{\sigma_1(\bX_R)}{\sigma_{\mathrm{min}}(\bX_{RC})} ||(\bV_R(\calJ,:)) ||_{2,\infty} \\
            & \overset{(c)}{\le} \frac{\sigma_1(\bX_R)}{\sigma_{\mathrm{min}}(\bX_{RC})} || \bV_R ||_{2,\infty} \\
            &\le \frac{\sigma_1(\bX)}{\sigma_{\mathrm{min}}(\bX_{RC})} || \bV_R ||_{2,\infty} \\
            & = \kappa(\bX) \frac{\sigma_{\mathrm{min}}(\bX)}{\sigma_{\mathrm{min}}(\bX_{RC})} || \bV_R ||_{2,\infty} \\
            & = \kappa(\bX) \frac{||\bX_{RC}^\dagger||_2}{||\bX^\dagger||_2} || \bV_R ||_{2,\infty}, 
        \end{aligned}
    \end{equation}
    where Lemma 4.2 in \cite{rCUR} gives (c). 
    
    Then, according to $ \bX_{RC} = \bU(\calI,:) \Sigma \bV(\calJ,:)^T$ and $||\Sigma^\dagger||_2 = \frac{1}{\sigma_{\mathrm{min}}(\bX)} = ||\bX^\dagger||_2$,
    we have 
    \begin{equation}
        \label{eq:XRCreverse}
        \begin{aligned}
            ||\bX_{RC}^\dagger||_2 &\le ||(\bV(\calJ,:)^T)^\dagger||_2 ||\Sigma^\dagger||_2 ||(\bU(\calI,:))^\dagger||_2 \\
            & \le ||(\bV(\calJ,:)^T)^\dagger||_2 ||\bX^\dagger||_2 ||(\bU(\calI,:))^\dagger||_2.
        \end{aligned}
    \end{equation}
    Thus, we substitute \cref{eq:XRCreverse} into \cref{eq:VRC2} to get
    \begin{align*}
        || \bV_{RC} ||_{2,\infty} & \le \kappa(\bX) ||(\bV(\calJ,:)^T)^\dagger||_2 ||(\bU(\calI,:))^\dagger||_2 || \bV_R ||_{2,\infty} \\
        & \le \kappa(\bX) \frac{1}{\sigma_{\mathrm{min}}(\bV(\calJ,:)) \sigma_{\mathrm{min}}(\bU(\calI,:))} || \bV_R ||_{2,\infty} \\
        & \overset{(d)}{\le} \frac{\kappa(\bX)}{1-\eta}  \sqrt{\frac{NT}{|\calI| |\calJ|}} || \bV_R ||_{2,\infty} \\
        &\le \frac{\kappa(\bX)}{1-\eta}  \sqrt{\frac{NT}{|\calI| |\calJ|}} \sqrt{\frac{\mu_2(\bX) r}{T}}
    \end{align*}
with probability at least $(1- p)^2$, where Lemma 3.4 in \cite{subsamp_sigma} gives (d).

\section{Proof of the \cref{thm_S}}
\label{proof:thm_S}

At last, we provide a lower bound on the number of samples (\cref{thm_S}), which depends only on the submatrix size and properties of $\bX$. We demonstrate that the sampling process, including the selection of rows/columns to obtain the submatrix and random sampling within $\bX_{RC}$, allows for successful reconstruction.

    According to the Cauchy-Schwartz inequality, we have
    $$ |(\bU_{RC} \bV_{RC}^T)(i,j)| \le ||\bU_{RC}(i,:)||_2 ||\bV_{RC}(j,:)||_2, $$
    and for all $(i,j)$
    \begin{equation}
        \label{eq:UV}
        ||\bU_{RC} \bV_{RC}^T||_\infty \le || \bU_{RC} ||_{2,\infty} || \bV_{RC} ||_{2,\infty}.
    \end{equation}
    Since $\calI$ and $\calJ$ are chosen uniformly from $\calV_\calG$ and $\calV_\calT$ based on \cref{proc:samp}, combining \cref{eq:UV} and \cref{lem:mu} we have
    \begin{align*}
        ||\bU_{RC} \bV_{RC}^T||_\infty &\le || \bU_{RC} ||_{2,\infty} || \bV_{RC} ||_{2,\infty} \\
        &\le \frac{(\kappa(\bX))^2}{(1-\eta)^{3/2}} \sqrt{\frac{\mu_1(\bX) \mu_2(\bX) Nr^2}{|\calI|} } \sqrt{\frac{r}{|\calI| |\calJ|}}
    \end{align*}
    for $\eta \in [0, 1)$ with probability at least $(1-p)^2$.

    Let $\mu_0 = \frac{(\kappa(\bX))^2}{(1-\eta)^{3/2}} \sqrt{\frac{\mu_1(\bX) \mu_2(\bX) Nr^2}{|\calI|} }$. The matrix $\bU_{RC} \bV_{RC}^T$ has a maximum entry bounded by $\mu_0 \sqrt{r / (|\calI| |\calJ|)}$. Let $n = \max{\{|\calI|,|\calJ|\}}$.
    Applying $\mu_0$ and $n$ to Theorem 2 in \cite{MC2}, we obtain that if 
    \begin{equation}
        \label{eq:S}
        \begin{aligned}
            |\calS| &\ge 32 \beta \mu_0^2 r (|\calI|+|\calJ|) \log^2(2n) \\
            &= \frac{32 \beta (\kappa(\bX))^4 r^2 N}{(1-\eta)^3} \mu_1(\bX) \mu_2(\bX) \frac{|\calI|+|\calJ|}{|\calI|} \log^2(2n)
        \end{aligned}
    \end{equation}
    for some $\beta > 1$, $\bX_{RC}$ can be reconstructed with probability
    $$ 1 - \frac{6 \log n}{(|\calI|+|\calJ|)^{2\beta-2}} - n^{2-2\beta^{1/2}}. $$

    Based on \cref{as_r}, we know that $|| \triangledown \bX(i,j) || < C$, so $|\bQ^T \bX(i,j)| < \infty$ and $|\bX \bD_1(i,j)| < \infty$. We define that 
    $$|| \triangledown \bX(i,j) ||_a = \sqrt{(\bQ^T \bX(i,j))^2 + (\bX \bD_1(i,j))^2 + a^2} $$
    for some positive parameter $a$, then 
    \begin{equation}
        \label{eq:delta1}
        0 < || \triangledown \bX(i,j) ||_a < \infty.
    \end{equation}
    Similarly, we have
    \begin{equation}
        \label{eq:delta2}
        \left| \bQ^T \bX(i,j) \right|_a = \sqrt{(\bQ^T \bX(i,j))^2 + a^2} < \infty
    \end{equation}
    and 
    \begin{equation}
        \label{eq:delta3}
        |\bX \bD_1(i,j)|_a = \sqrt{(\bX \bD_1(i,j))^2 + a^2} < \infty.
    \end{equation}

    According to Eq. (5.7) in \cite{XRC2X}, we have
    \begin{align*}
        \triangledown & \left( \frac{\triangledown \bX(i,j)}{|| \triangledown \bX(i,j) ||_a} \right) \\
        & = \frac{1}{|| \triangledown \bX(i,j) ||_a^3} ( |\bX \bD_1(i,j)|_a^2 \bX(:,j)^T \bL \bX(:,j) \\
        & \quad + \left| \bQ^T \bX(i,j) \right|_a^2 (\bX \bD_2)(i,j) \\
        & \quad - 2 \bQ^T \bX(i,j) \bX \bD_1(i,j)  \bQ^T \bX \bD_1(i,j) ). 
    \end{align*}
    Then, combining \cref{as_r} and \cref{eq:delta1,eq:delta2,eq:delta3} we have 
    $$\left| \triangledown \left( \frac{\triangledown \bX(i,j)}{|| \triangledown \bX(i,j) ||_a} \right) \right| < \infty.$$
    Thus, we can reconstruct $\bX$ from $\bX_{RC}$ by solving the TV inpainting model according to Theorem 4.4 in \cite{euler}. 
    
    As a result, if \cref{eq:S} holds, we are able to reconstruct $\bX$ from $\calS$ with probability at least 
    $$ (1-\delta)^2 (1-p)^2 -\frac{6 \log{n}}{(|\calI|+|\calJ|)^{2\beta-2}} -n^{2-2\beta^{1/2}}. $$

\section{Algorithm for opt. prob. based on \cref{eq:optm}}
\label{sec:algo}

In \cref{eq:optm}, the signal spectral coefficients are constrained using the $\ell_1$-norm. Unlike the non-convex $\ell_0$-norm, the $\ell_1$-norm is convex, making the optimization problem easier to solve. However, since the $\ell_1$-norm is defined as the sum of the absolute values of non-zero elements, it imposes sparsity constraints but also affects the magnitude of these elements. To overcome the influence of the signal magnitudes, a reweighting scheme is introduced \cite{reweighted,reweighted_l1}. 

To enhance robustness and accelerate the solution of the optimization problem, we incorporate the error matrix $\bE$ into the constraints \cite{RPCA}.
As a result, the optimization problem based on \cref{eq:optm} is 
\begin{equation}
    \begin{aligned}
        &\min_{\hat{\bX}, \bE} \quad f_{\text{obj}} = g(\lambda_i (\hat{\bX}^T \hat{\bX})) + \gamma_\calG || \bW_\calG \odot \bF_\calG||_1\\
        & \qquad \qquad \quad \quad + \gamma_\calT || \bW_\calT \odot \bF_\calT||_1 + \gamma_d ||\hat{\bX} \bD_2||_F^2 \\
        & \begin{array}{c@{\quad \quad \quad}l@{\quad}l}
        s.t. & \hat{\bX} = \Psi_\calG \bF_\calG, \hat{\bX}^T = \Psi_\calT \bF_\calT, \\
             & \bX_\calS = \hat{\bX} + \bE, \scP_\calS(\bE) = 0,
        \end{array}
    \end{aligned}
\label{eq:optm1}
\end{equation}
where $\bE$ is an auxiliary error matrix and $\scP_\calS(\cdot)$ be the projection operator onto $\calS$, 
\begin{equation*}
    \scP_\calS(\bE)(i,j) = \left \{ 
        \begin{array}{ll}
        \bE(i,j), & {\text{if $(i,j) \in \calS$};} \\ 
        0, & {\text{otherwise}.} \end{array}\right.
\end{equation*}

The LSSP algorithm consists of three layers of loops. The outermost loop is the reweighting framework, with $p$ denoting the iteration index. For example, the variable $\bF_\calG$ at the $p$-th iteration is represented as $\bF_\calG^{\{p\}}$. The middle loop updates each variable, with $k$ denoting the iteration index, and the result of the $k$-th iteration is $\bF_\calG^{(k)}$. During each iteration of the middle loop, the updates of $\bF_\calG$ and $\bF_\calT$ are required to solve the optimization problems \cref{eq:optm_FG} and \cref{eq:optm_FT}, leading to the innermost loop. In this loop, indexed by $q$, $\bF_\calG^{(k),q}$ represents the $q$-th iteration of $\bF_\calG$ during the $k$-th variable update.

The reweighting scheme is an effective iterative method that significantly enhances signal reconstruction by reinforcing the sparsity of the coefficients. The key idea behind reweighted $\ell_1$ norm minimization is to apply large weights to suppress zero or near-zero elements and small weights to promote non-zero entries. As suggested in \cite{reweighted_l1}, the weights in reweighted $\ell_1$ norm minimization are inversely proportional to the magnitude of the coefficients. The weight matrix is updated as follows
\begin{equation}
    \left \{ 
        \begin{array}{ll}
        \bW_\calG^{\{p+1\}}(i, j) = \frac{1}{|\bF_\calG^{\{p\}}(i,j)| + \zeta}  \\ 
        \bW_\calT^{\{p+1\}}(i, j) = \frac{1}{|\bF_\calT^{\{p\}}(i,j)| + \zeta} 
        \end{array}\right.
\label{eq:Wp+1}
\end{equation}
for parameter $\zeta > 0$.

In the $p$-th iteration, given the weight matrices $\bW_\calG^{\{p\}}$ and $\bW_\calT^{\{p\}}$, we solve \cref{eq:optm} subject to the equation constraints.

The non-convex regularizer is optimized using an iteratively reweighted least squares method \cite{NCMC}. In each iteration $k$, A weight matrix $\bR^{(k)}$ is updated based on $\hat{\bX}^{(k-1)}$, and then $\hat{\bX}^{(k)}$ is updated according to $\bR^{(k)}$ . 

By applying the augmented Lagrangian method, we transform \cref{eq:optm} into an unconstrained optimization problem, where the Lagrangian function is expressed as:
\begin{equation}
    \begin{aligned}
        & L(\hat{\bX}, \bF_\calG, \bF_\calT, \bE, \bY_1, \bY_2, \bY_3) \\
        & = \text{Tr}(\bR^T h(\hat{\bX})) + \gamma_d ||\hat{\bX} \bD_2||_F^2 \\
        & \quad + \gamma_\calG || \bW_\calG \odot \bF_\calG||_1 + \gamma_\calT || \bW_\calT \odot \bF_\calT||_1 \\
        & \quad + \langle \bY_1, \hat{\bX} - \Psi_\calG \bF_\calG \rangle + \langle \bY_2, \hat{\bX}^T - \Psi_\calT \bF_\calT \rangle \\
        & \quad + \frac{\mu_1}{2} ||\hat{\bX} - \Psi_\calG \bF_\calG||_F^2 + \frac{\mu_2}{2} ||\hat{\bX}^T - \Psi_\calT \bF_\calT||_F^2 \\
        & \quad + \langle \bY_3, \bX_\calS - \hat{\bX} - \bE \rangle + \frac{\mu_3}{2}|| \bX_\calS - \hat{\bX} - \bE||_F^2,
    \end{aligned}
\label{eq:Lagrangian_0}
\end{equation}
where $\bY_1$, $\bY_2$ and $\bY_3$ are the Lagrange multiplier matrices, $\mu_1$, $\mu_2$ and $\mu_3$ are the penalty factors, and $\bX_\calS$ is the incomplete observed matrix.

To efficiently solve \cref{eq:Lagrangian_0} with multiple unknown variables, we adopt the alternating direction method of multipliers (ADMM), optimizing each variable while keeping the others fixed.

Optimizing $\bF_\calG$ means solving the following optimization problem:
\begin{equation}
    \begin{aligned}
        \bF_\calG^{(k+1)} = & \arg \min_{\bF_\calG} \gamma_\calG || \bW_\calG \odot \bF_\calG||_1 + \langle \bY_1^{(k)}, \hat{\bX}^{(k)} - \Psi_\calG \bF_\calG \rangle \\
        &+ \frac{\mu_1^{(k)}}{2} ||\hat{\bX}^{(k)} - \Psi_\calG \bF_\calG||_F^2.
    \end{aligned}
\label{eq:optm_FG}
\end{equation}
Instead of solving \cref{eq:optm_FG} directly, we estimate $\bF_\calG$ using the following approximation model. Let $f_1(\bF_\calG) = \langle \bY_1^{(k)}, \hat{\bX}^{(k)} - \Psi_\calG \bF_\calG \rangle + \frac{\mu_1^{(k)}}{2} ||\hat{\bX}^{(k)} - \Psi_\calG \bF_\calG||_F^2$. By introducing a proximal variable $\bA_1$ \cite{FISTA}, we have
\begin{equation}
    \begin{aligned}
        \text{prox}_{\tau_1} (\bA_1) =& \arg \min_{\bF_\calG} \gamma_\calG || \bW_\calG \odot \bF_\calG||_1 \\
        &+ \frac{1}{2 \tau_1} ||\bF_\calG - (\bA_1 - \tau_1 \triangledown f_1(\bA_1)) ||_2^2
    \end{aligned}
\end{equation}
where $\tau_1 >0 $ is a constant, and 
$$\triangledown f_1(\bA_1) = \mu_1^{(k)} \bA_1 - \Psi_\calG^T \bY_1^{(k)} - \mu_1^{(k)} \Psi_\calG^T \hat{\bX}^{(k)}. $$
Now we can solve the $\text{prox}_{\tau_1} (\bA_1)$ by nonuniform soft thresholding on $\bA_1 - \tau_1 \triangledown f_1(\bA_1)$ \cite{reweighted}, \textit{i.e.},
\begin{equation}
    \bF_\calG^{(k),q+1} = \text{soft}_{\frac{\gamma_\calG \bW_\calG}{\mu_1^{(k)}}} (\bA_1^q - \tau_1 \triangledown f_1(\bA_1^q)),
\label{eq:FGkq+1}
\end{equation}
where 
$$\text{soft}_\bW (\bA) = \{ \text{sign}(\bA(i,j)) \max(|\bA(i,j)|-\bW(i,j), 0) \}$$ 
is the shrinkage operator.
Then, the proximal variable $\bA$ can be update by
\begin{equation}
    \left \{ 
        \begin{array}{l}
        b_1^{q+1} = \frac{1+\sqrt{4 (b_1^q)^2+1}}{2},  \\ 
        \bA_1^{q+1} = \bF_\calG^{(k),q+1} \! + \frac{b_1^q - 1}{b_1^{q+1}} (\bF_\calG^{(k),q+1} \! - \bF_\calG^{(k),q}),
        \end{array}\right.
\label{eq:FGkA}
\end{equation}
where $b_1^q$ is a positive parameter with $b_1^1 = 1$. The solution to this proximal problem is $\bF_\calG^{(k+1)}$, \textit{i.e.}, $\bF_\calG^{(k+1)} = \bF_\calG^{(k),Q}$

Consider $\bF_\calT$, the optimization problem is
\begin{equation}
    \bF_\calT^{(k+1)} = \arg \min_{\bF_\calT} \gamma_\calT || \bW_\calT \odot \bF_\calT||_1 + f_2(\bF_\calT),
\label{eq:optm_FT}
\end{equation}
where $f_2(\bF_\calT) = \langle \bY_2^{(k)}, (\hat{\bX}^{(k)})^T - \Psi_\calT \bF_\calT \rangle + \frac{\mu_2^{(k)}}{2} ||(\hat{\bX}^{(k)})^T - \Psi_\calT \bF_\calT||_F^2$. 
In like manner, $\bF_\calT^{(k+1)}$ can be approximated as follows:
\begin{equation}
    \left \{ 
        \begin{array}{l}
        \bF_\calT^{(k),q+1} = \text{soft}_{\frac{\gamma_\calT \bW_\calT}{\mu_2^{(k)}}} (\bA_2^q - \tau_2 \triangledown f_2(\bA_2^q)),  \\ 
        b_2^{q+1} = \frac{1+\sqrt{4 (b_2^q)^2+1}}{2},  \\
        \bA_2^{q+1} = \bF_\calT^{(k),q+1} \! + \frac{b_2^q - 1}{b_2^{q+1}} (\bF_\calT^{(k),q+1} \! - \bF_\calT^{(k),q}),
        \end{array}\right.
\label{eq:FTkq+1}
\end{equation}
where $\bA_2$ is a proximal variable, $\tau_2 > 0$ is a constant, $b_2^q$ is a positive parameter with $b_2^1 = 1$, and 
$$ \triangledown f_2(\bA_2^q) = \mu_2^{(k)} \bA_2 - \Psi_\calT^T \bY_2^{(k)} - \mu_2^{(k)} \Psi_\calT^T (\hat{\bX}^{(k)})^T. $$
Eventually, we get $\bF_\calT^{(k+1)} = \bF_\calT^{(k),Q}$.

To solve $\hat{\bX}$, we first eigenvalue decompose $h(\hat{\bX})$ as
$$ h(\hat{\bX}^{(k)}) = \bU_h^{(k)} \Lambda_h^{(k)} (\bU_h^{(k)})^T. $$
Then $\bR$ is updated by
$$\bR^{(k+1)} = \bU_h^{(k)} g'(\Lambda_h^{(k)}) (\bU_h^{(k)})^T, $$
where $g'(x) = \frac{1}{2\sqrt{x}(\sqrt{x}+1)}$.
Let 
\begin{equation*}
    \begin{aligned}
        & L_1(\hat{\bX}, \bF_\calG^{(k+1)}, \bF_\calT^{(k+1)}, \bE^{(k)}, \bY_1^{(k)}, \bY_2^{(k)}, \bY_3^{(k)}) \\
        & = \text{Tr}((\bR^{(k+1)})^T h(\hat{\bX})) + \gamma_d ||\hat{\bX} \bD_2||_F^2 \\
        & \quad + \langle \bY_1^{(k)}, \hat{\bX} - \Psi_\calG \bF_\calG^{(k+1)} \rangle + \langle \bY_2^{(k)}, \hat{\bX}^T - \Psi_\calT \bF_\calT^{(k+1)} \rangle \\
        & \quad + \frac{\mu_1^{(k)}}{2} ||\hat{\bX} - \Psi_\calG \bF_\calG^{(k+1)}||_F^2 + \frac{\mu_2^{(k)}}{2} ||\hat{\bX}^T - \Psi_\calT \bF_\calT^{(k+1)}||_F^2 \\
        & \quad + \langle \bY_3^{(k)}, \bX_\calS - \hat{\bX} - \bE^{(k)} \rangle + \frac{\mu_3^{(k)}}{2}|| \bX_\calS - \hat{\bX} - \bE^{(k)}||_F^2,
    \end{aligned}
\end{equation*}
By taking the derivative with respect to $\hat{\bX}$ and setting it to zero, we have 
\begin{equation*}
    \begin{aligned}
        \frac{\partial L_1}{\partial \hat{\bX}} = & 2 \hat{\bX} \bR^{(k+1)} + 2 \gamma_d \hat{\bX} \bD_2 \bD_2^T + \bY_1^{(k)} + (\bY_2^{(k)})^T \\
        & + \mu_1^{(k)} (\hat{\bX} - \Psi_\calG \bF_\calG^{(k+1)}) + \mu_2^{(k)} (\hat{\bX} - (\bF_\calT^{(k+1)})^T \Psi_\calT^T ) \\
        & - \bY_3^{(k)} + \mu_3^{(k)} (\hat{\bX} - \bX_\calS + \bE^{(k)}) = 0. 
    \end{aligned}
\end{equation*}
The optimal solution of $\hat{\bX}$ can be obtained by
\begin{equation}
    \begin{aligned}
        &\hat{\bX}^{(k+1)} \\
        &= ( \bY_3^{(k)} - \bY_1^{(k)} - (\bY_2^{(k)})^T + \mu_1^{(k)} \Psi_\calG \bF_\calG^{(k+1)} \\
        & \quad + \mu_2^{(k)} (\bF_\calT^{(k+1)})^T \Psi_\calT^T + \mu_3^{(k)} (\bX_\calS - \bE^{(k)}) ) \\
        & \quad (2 \bR^{(k+1)} + 2 \gamma_d \bD_2 \bD_2^T + (\mu_1^{(k)}+\mu_2^{(k)}+\mu_3^{(k)})\bI )^\dagger
    \end{aligned}
\label{eq:hatXk+1}
\end{equation}

To solve $\bE$, the Lagrangian function becomes
\begin{equation*}
    \begin{aligned}
        & L_2(\hat{\bX}^{(k+1)}, \bE, \bY_3^{(k)}) \\
        & = \langle \bY_3^{(k)}, \bX_\calS - \hat{\bX}^{(k+1)} - \bE \rangle + \frac{\mu_3^{(k)}}{2}|| \bX_\calS - \hat{\bX}^{(k+1)} - \bE||_F^2,
    \end{aligned}
\end{equation*}
In the same way, we take the derivative with respect to $\bE$ and set it to zero
\begin{equation*}
    \frac{\partial L_2}{\partial \bE} = -\bY_3^{(k)} + \mu_3^{(k)}(\bE - \bX_\calS + \hat{\bX}^{(k+1)}) = 0.
\end{equation*}
When updating the error matrix $\bE$, we only need to update the entries in $\calS^c$ (the complement of $\calS$). Thus, we have 
\begin{equation}
    \bE^{(k+1)} = \textbf{0} + \scP_{\calS^c} \left( \frac{1}{\mu_3^{(k)}} \bY_3^{(k)} + \bX_\calS - \hat{\bX}^{(k+1)} \right),
\label{eq:Ek+1}
\end{equation}
where $\textbf{0}$ is an all-zero matrix.

Furthermore, the Lagrange multiplier matrices and penalty factors can be updated by
\begin{equation}
    \left \{ 
        \begin{array}{l}
        \bY_1^{(k+1)} = \bY_1^{(k)} + \mu_1^{(k)}(\hat{\bX}^{(k+1)} - \Psi_\calG \bF_\calG^{(k+1)}),  \\ 
        \bY_2^{(k+1)} = \bY_2^{(k)} + \mu_2^{(k)}((\hat{\bX}^{(k+1)})^T - \Psi_\calT \bF_\calT^{(k+1)}),  \\
        \bY_3^{(k+1)} = \bY_3^{(k)} + \mu_3^{(k)}(\bX_\calS - \hat{\bX}^{(k+1)} - \bE^{(k+1)}),  \\
        \mu_1^{(k+1)} = \rho_1 \mu_1^{(k)}, \\
        \mu_2^{(k+1)} = \rho_2 \mu_2^{(k)}, \\
        \mu_3^{(k+1)} = \rho_3 \mu_3^{(k)},
        \end{array}\right.
\label{eq:Ymuk+1}
\end{equation}
where $\rho_1 > 1$, $\rho_2 > 1$, and $\rho_3 >1$ are constants that ensure the increase of the penalty factors.

\bibliographystyle{IEEEtran}
\bibliography{IEEEabrv,references}

\begin{IEEEbiography}[{\includegraphics[width=1in,height=1.25in,clip,keepaspectratio]{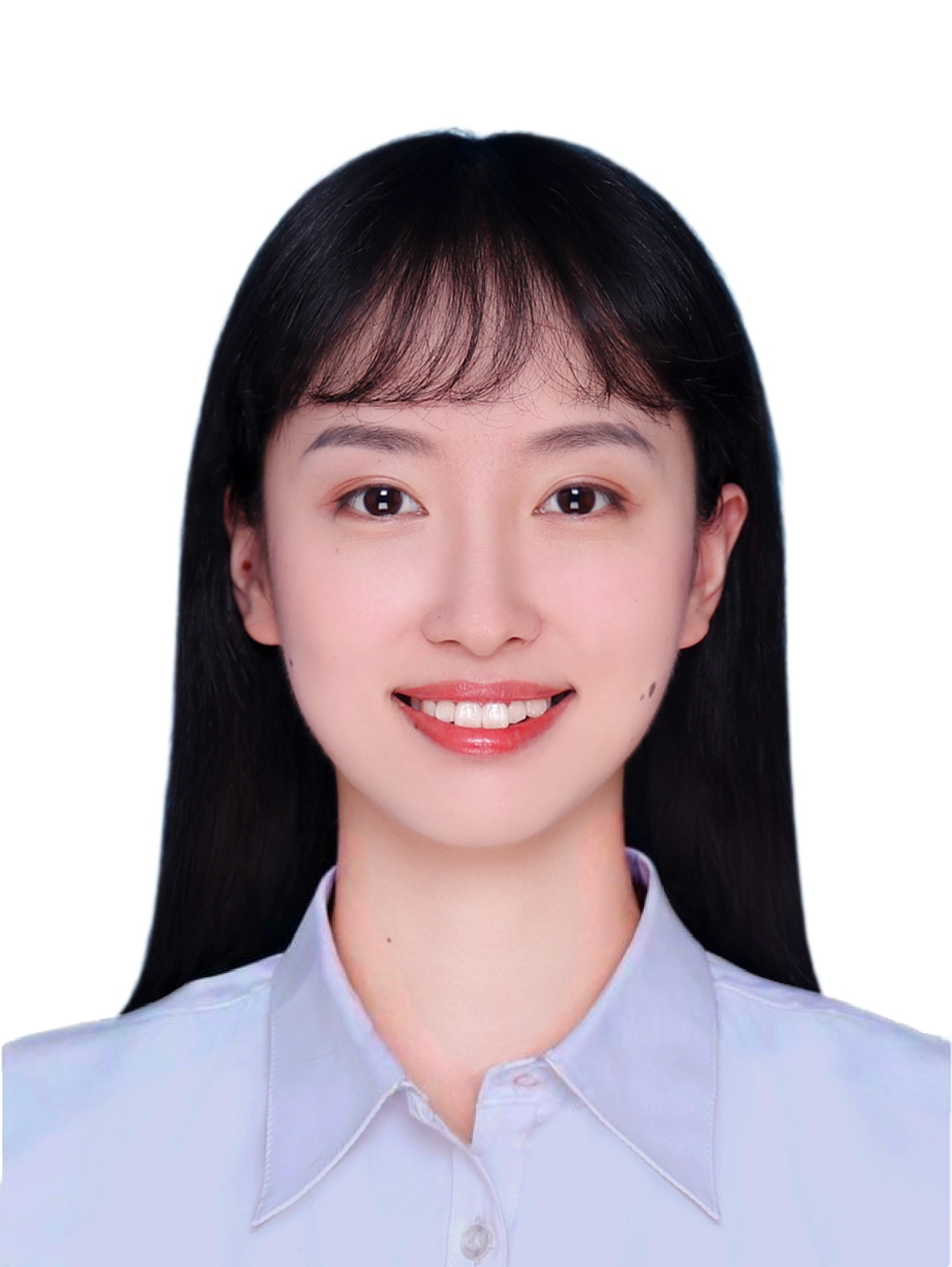}}]{Hang Sheng}
(Student Member, IEEE) received the B.Sc. degree from the College of Engineering, Nanjing Agricultural University, China, in 2020. She is currently working toward the Ph.D. degree with College of Future Information Technology, Fudan University, China. Her research interests include graph signal processing and machine learning.
\end{IEEEbiography}

\begin{IEEEbiography}[{\includegraphics[width=1in,height=1.25in,clip,keepaspectratio]{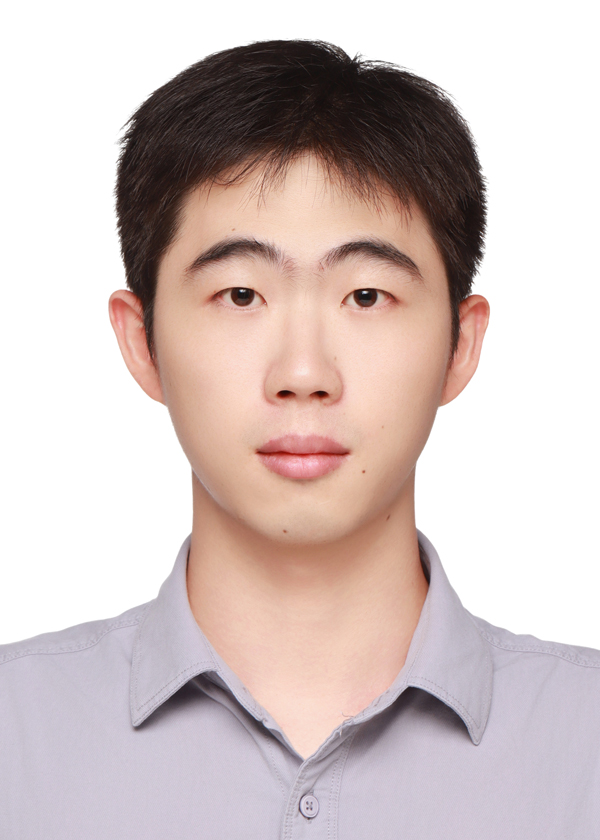}}]{Qinji Shu}
received the B.Sc. degree in electronic engineering from Fudan University, Shanghai, China, in 2022. He is currently a Master's student with College of Future Information Technology, Fudan University. His research focuses on the theoretical transferability in graph-based learning.
\end{IEEEbiography}

\begin{IEEEbiography}[{\includegraphics[width=1in,height=1.25in,clip,keepaspectratio]{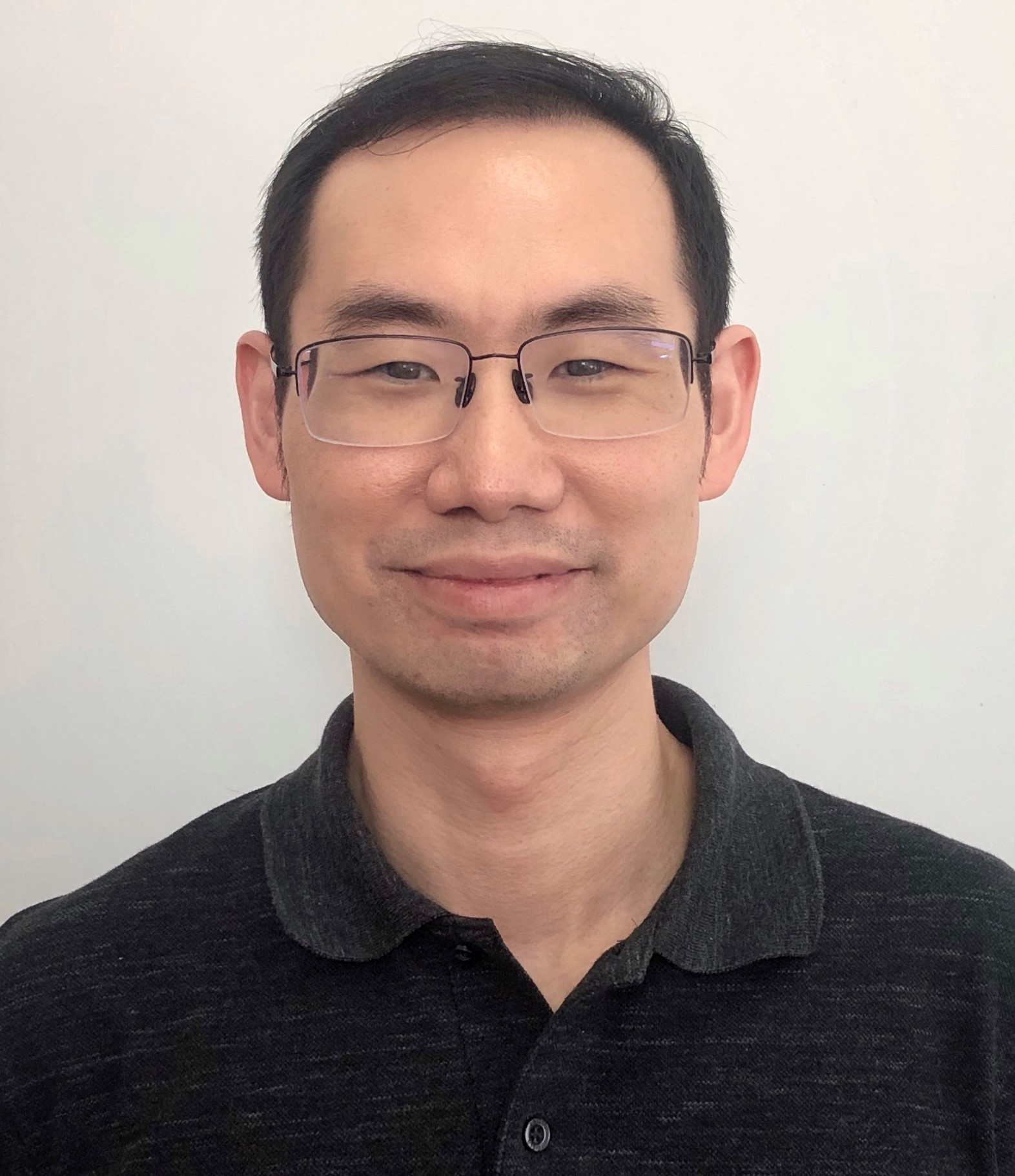}}]{Hui Feng}
(Member, IEEE) received the B.Sc., M.Sc., and Ph.D. degrees in electronic engineering from Fudan University, Shanghai, China, in 2003, 2006, and 2014, respectively. He is currently an Associate Professor with College of Future Information Technology, Fudan University. His research focuses on intelligent signal processing and electronic system design.
\end{IEEEbiography}

\begin{IEEEbiography}[{\includegraphics[width=1in,height=1.25in,clip,keepaspectratio]{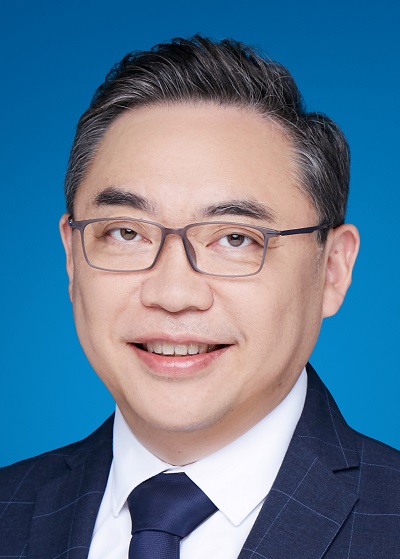}}]{Bo Hu}
(Member, IEEE) received the B.S. and Ph.D. degrees in electronic engineering from Fudan University, Shanghai, China, in 1990 and 1996, respectively. He is currently a Professor with College of Future Information Technology, Fudan University. His research interests include intelligent image processing, digital communication, and intelligent system design.
\end{IEEEbiography}

\end{document}